\documentclass[journal]{IEEEtran}

\usepackage{color}
\usepackage{graphicx}
\usepackage{mathtools}
\usepackage{placeins}
\usepackage[utf8]{inputenc}
\usepackage{bm} 

\usepackage{tikz}
\usetikzlibrary{positioning}
\usetikzlibrary{shapes}
\usetikzlibrary{spy}
\tikzset{>=latex}
\pgfdeclarelayer{bg}
\pgfsetlayers{bg,main}

\usepackage{pgfplots}
\pgfplotsset{compat=1.14}

\usepackage{cite}

\makeatletter
\def\ps@IEEEtitlepagestyle{
	\def\@oddfoot{\mycopyrightnotice}
	\def\@evenfoot{}
}
\def\mycopyrightnotice{
	{\scriptsize
	\begin{minipage}{\textwidth}
	\centering
	Coypright~\copyright~2018 IEEE.
	Personal use of this material is permitted.
	Permission from IEEE must be obtained for all other uses, in any current or future media, including reprinting/republishing this material for advertising or promotional purposes, creating new collective works, for resale or redistribution to servers or lists, or reuse of any copyrighted component of this work in other works.
	\end{minipage}
	}
}

\begin{document}

\title{High-Precision Measurement of Sine and Pulse Reference Signals using Software-Defined Radio}

\author{\IEEEauthorblockN{
Carsten Andrich\IEEEauthorrefmark{1},   
Alexander Ihlow\IEEEauthorrefmark{2},   
Julia Bauer\IEEEauthorrefmark{1},    
Niklas Beuster\IEEEauthorrefmark{2},
Giovanni Del Galdo\IEEEauthorrefmark{1}\IEEEauthorrefmark{2} 
} \\
\IEEEauthorblockA{\IEEEauthorrefmark{1}Fraunhofer Institute for Integrated Circuits IIS, 98693 Ilmenau, Germany} \\
\IEEEauthorblockA{\IEEEauthorrefmark{2}Institute for Information Technology, Technische Universit\"at Ilmenau, 98693 Ilmenau, Germany} \\
\IEEEauthorblockA{Phone: +49 3677 69-4281, Email: carsten \textit{dot} andrich \textit{at} iis \textit{dot} fraunhofer \textit{dot} de}

\thanks{Published in IEEE Transactions on Instrumentation and Measurement, vol.~67, no.~5, pp.~1132 -- 1141, May 2018. DOI: 10.1109/TIM.2018.2794940}
}

\maketitle

\begin{abstract}
This paper addresses simultaneous, high-precision measurement and analysis of generic reference signals by using inexpensive commercial off-the-shelf Software Defined Radio hardware.
Sine reference signals are digitally down-converted to baseband for the analysis of phase deviations.
Hereby, we compare the precision of the fixed-point hardware Digital Signal Processing chain with a custom Single Instruction Multiple Data (SIMD) x86 floating-point implementation.
Pulse reference signals are analyzed by a software trigger that precisely locates the time where the slope passes a certain threshold.
The measurement system is implemented and verified using the Universal Software Radio Peripheral (USRP) N210 by Ettus Research LLC.
Applying standard 10\,MHz and 1\,PPS reference signals for testing, a measurement precision (standard deviation) of 0.36\,ps and 16.6\,ps is obtained, respectively.
In connection with standard PC hardware, the system allows long-term acquisition and storage of measurement data over several weeks.
A comparison is given to the Dual Mixer Time Difference (DMTD) and Time Interval Counter (TIC), which are state-of-the-art measurement methods for sine and pulse signal analysis, respectively.
Furthermore, we show that our proposed USRP-based approach outperforms measurements with a high-grade Digital Sampling Oscilloscope.
\end{abstract}

\begin{IEEEkeywords}
Phase Measurement, Time Measurement, Digital Signal Processing, Measurement Techniques, Software Defined Radio, Computerized Instrumentation, Reference Signal, 10\,MHz, 1 Pulse Per Second (1\,PPS), Time Domain Analysis, Time Interval Counter (TIC), Dual Mixer Time Difference (DMTD).
\end{IEEEkeywords}

\IEEEpeerreviewmaketitle

\section{Introduction}

\IEEEPARstart{R}{eference} signals are relied upon, when electronic measurements are required to meet demanding timing or frequency precision constraints.
Either a single device can be coupled with a more accurate external reference, or multiple devices can be synchronized using an appropriately distributed reference signal.
A particularly challenging example is the synchronization of a spatially distributed measurement setup, necessitating the use of multiple references, one for each location of the measurement.

Exemplary frequency standards used to generate highly precise reference signals are atomic clocks or GPS Disciplined Oscillators (GPSDO).
These standards usually output both a sine wave and a pulse signal, while lab-grade measurement devices have inputs for both signal types.

\subsection{Sine Wave Reference Signal}

A simplified sine wave reference signal model is given by Equation~\ref{eqn:x_sine}, consisting only of a perfect sine wave at the reference frequency $ f_\mathrm{r} $ and a time-variant phase error term~$ \varphi_\mathrm{e}(t) $, neglecting signal levels and noise \cite{riley2008handbook}.
\begin{equation}
\label{eqn:x_sine}
x_\mathrm{sine}(t) = \mathrm{sin}(2 \pi f_\mathrm{r} t + \varphi_\mathrm{e}(t))
\end{equation}

Sine wave reference signals are used for \textit{relative} time synchronization, meaning that no drift between the time base of the reference signal and the time base of the synchronized system will occur.
This is usually achieved by using a Phase Locked Loop (PLL) to lock the internal clock of the measurement system to the reference signal.
In case of Radio Frequency (RF) measurement devices, local oscillators may also be locked to the reference signal to improve their frequency accuracy or to enable RF phase coherence.

The reference frequency $ f_\mathrm{r} $ may depend on the intended application, but 10\,MHz signals enjoy widespread device support, with 5\,MHz and 100\,MHz being alternative choices.
Although most applications rely on sine waves, some may use rectangular signals instead.
Fortunately, all considerations for sine waves apply to rectangular signals as well, since the latter can be transformed into the former by means of a band-pass filter.

Due to their periodic nature, sine (and rectangular) reference signals cannot provide an absolute time base.
For this purpose, a trigger event is needed, i.e. a Pulse Reference Signal.

\subsection{Pulse Reference Signal}
\label{sec:pulse_sig}

A simplified pulse signal model is given by Equation~\ref{eqn:x_pulse}, consisting of a pulse waveform $ \mathrm{pls}(t) $ convolved with an infinite series of Dirac pulses $ \delta(t) $, which are equally spaced $ T_\text{period} $ apart, except for a time error $ t_\mathrm{e}[n] $.
\begin{equation}
\label{eqn:x_pulse}
x_\mathrm{pulse}(t) = \mathrm{pls}(t) \ast \!\!\!\! \sum_{n = - \infty}^{\infty} \!\!\! \delta(t - n \cdot T_\text{period} - t_\mathrm{e}[n])
\end{equation}

The signal's timing accuracy depends on its rise-time \cite{siccardi2016delay}, which is inversely proportional to its bandwidth.
As signal bandwidth is always finite and free of discontinuities (i.e. no rectangular band-limitation), a simple $ \mathrm{pls}(t) $ signal model is given by the $ 2^\mathrm{nd} $~order low-pass characteristic.
This model may be an oversimplification, but it shows two typical characteristics of pulse signals, namely a finite rise-time and transient overshoots.

We will model a single edge of $ \mathrm{pls}(t) $ using the $ 2^\mathrm{nd} $~order low-pass step response $ h_{\mathrm{LP}2}(t) $, with the angular cutoff frequency $ \omega_0 $ and the damping coefficient $ \zeta $:
\[
h_{\mathrm{LP}2}(t) = 1 - \dfrac{1}{\sqrt{1-\zeta^2}}
     \cdot \mathrm{e}^{-\zeta \omega_0 t}
     \cdot \mathrm{sin} \! \left( \sqrt{1-\zeta^2} \, \omega_0 t + \mathrm{acos}(\zeta) \right)
\]

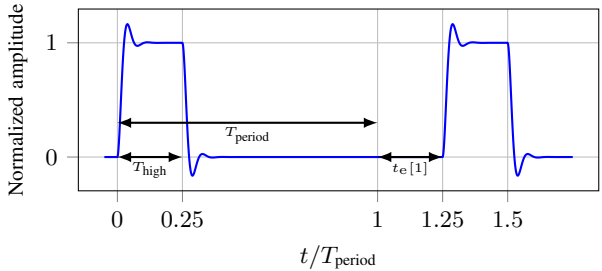
\begin{figure}[ht!]
\begin{center}
\vspace{-5mm}
\begin{tikzpicture}
\tikzstyle{every node}=[font=\small]

\begin{axis}[
		width=8.5cm,
		height=4.0cm,
		xmin=-0.15,
		xmax=1.85,
		grid=both,
		tick align=outside,
		xtick pos=left,
		ytick pos=bottom,
		xtick={0, 0.25, 1, 1.25, 1.5},
		ytick={-1, 0, 1},
		xlabel=$ t / T_\text{period} $,
		ylabel=\footnotesize Normalized amplitude
	]
	\addplot[thick,blue] table {pulse.csv};
	
	\node[anchor=north,yshift=0.4ex] at (axis cs:0.125,0){\tiny $ T_\text{high} $};
	\draw[<->,thick](axis cs:0,0)--(axis cs:0.25,0);
	
	\node[anchor=north,yshift=0.4ex] at (axis cs:0.5,0.3){\tiny $ T_\text{period} $};
	\draw[<->,thick](axis cs:0,0.3)--(axis cs:1,0.3);
	
	\node[anchor=north,yshift=0.4ex] at (axis cs:1.125,0){\tiny $ t_\mathrm{e}[1] $};
	\draw[<->,thick](axis cs:1,0)--(axis cs:1.25,0);
\end{axis}
\end{tikzpicture}
\vspace{-5mm}
\end{center}
\caption{Exemplary pulse signal modeled according to Equation~\ref{eqn:x_pulse}.
Note that every pulse is offset by an individual time-error $ t_\mathrm{e}[n] $.}
\label{plot:pulse}
\end{figure}

Based on said step response, a single pulse with high-level duration $ T_\text{high} $ will be modeled as follows, again disregarding signal amplitude (see Figure~\ref{plot:pulse} for an example):
\[
\mathrm{pls}(t) = 
\begin{cases}
	0, & \text{if} \; t < 0 \\
	h_{\mathrm{LP}2}(t), & \text{if} \; 0 \leq t < T_\text{high} \\
	h_{\mathrm{LP}2}(t) - h_{\mathrm{LP}2}(t - T_\text{high}), & \text{if} \; t \geq T_\text{high} \\
\end{cases}
\]

Pulse signals complement sine wave reference signals by enabling \textit{absolute} time synchronization.
Typically, a so called 1 Pulse Per Second (1\,PPS) signal with a $ T_\text{period} $ of 1 second is used.
For example, when using a GPSDO as a 1\,PPS source, the rising edge of the signal coincides with the Coordinated Universal Time (UTC) second.
When a measurement device is accurately synchronized to full seconds using a 1\,PPS signal, less precise mechanism, e.g. the Network Time Protocol (NTP) or a GPSDO's serial output, can be used to achieve absolute synchronization to UTC.

\subsection{Reference Signal Measurement}

Reference signal generators are purpose-built to provide highly stable signals and thus contain high-grade oscillators. These easily outperform those of laboratory measurement equipment and attempting to judge the stability of a reference signal using a measurement device with an inferior clock would defeat the purpose of the measurement.
Therefore, the quality of reference signals is gauged by directly comparing two reference signals, eliminating the need for a highly-precise measurement clock.

To compare two sine signals of the same (or potentially different) frequency $ x_{\text{sine},1}(t) $ and $ x_{\text{sine},2}(t) $ (cf. Equation~\ref{eqn:x_sine}), the difference between their phase errors $ \Delta\varphi(t) $ must be measured:
\[
\Delta\varphi(t) = \varphi_{\mathrm{e},1}(t) - \varphi_{\mathrm{e},2}(t)
\]
From this, the associated time error $ \Delta t_\text{sine}(t) $ can be computed:
\begin{equation}
\label{eqn:t_sine}
\Delta t_\text{sine}(t) = \dfrac{\Delta\varphi(t)}{2 \pi f_\mathrm{r}}
\end{equation}

The comparison of two pulse signals $ x_{\text{pulse},1}(t) $ and $ x_{\text{pulse},2}(t) $ (cf. Equation~\ref{eqn:x_pulse}) requires to determine the difference between their respective time errors $ \Delta t_\text{pulse}[n] $:
\[
\Delta t_\text{pulse}[n] = t_{\mathrm{e},1}[n] - t_{\mathrm{e},2}[n]
\]

As \textit{absolute} time synchronization requires both signal types, a reference signal measurement system should compare both signal types of two reference sources simultaneously.

\FloatBarrier

\section{State-of-the-art of Measurement Systems}

\IEEEPARstart{T}{wo} reference signal measurement architectures currently enjoy widespread use.
The Dual Mixer Time Difference (DMTD) method is used for sinusoidal signals and the Time Interval Counter (TIC) technique is employed for pulse signals.

\subsection{Time Interval Counter}
\label{sec:tic}

A TIC is a digital circuit used to measure the time difference~$ \Delta t_\text{pulse}[n] $ between two (or more) pulse signals' rising edges.
A basic implementation relies on a high-frequency digital clock, which is gated by digital circuitry responsible for pulse detection.
The clock cycles between rising edges are counted to determine $ \Delta t_\text{pulse}[n] $.
Therefore, a TIC's resolution and accuracy depends on its digital clock.
\cite{kalisz2004review}

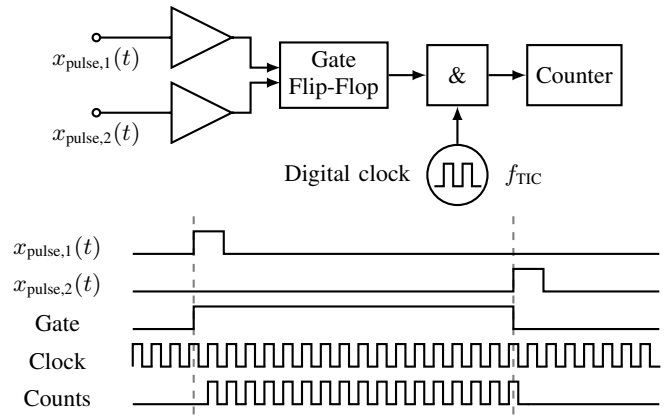
\begin{figure}[ht!]
\begin{center}
\vspace{-2mm}
\begin{tikzpicture}
\tikzstyle{every node}=[font=\small]
\tikzstyle{every path}=[thick]

\tikzset{block/.style={%
	rectangle, draw, minimum height=8mm, minimum width=8mm
}}
\tikzset{amp/.style={
	rectangle, minimum width=8mm+\pgflinewidth, minimum height=8mm+\pgflinewidth, inner sep=0,
	path picture={%
		\draw[-] (-4mm,-4mm) -- ++(0mm,8mm) -- ++(8mm,-4mm) -- cycle;
	}
}}
\tikzset{clk/.style={
	circle, draw, minimum size=8mm, inner sep=0,
	path picture={%
		\draw (-0.30625,-0.15) -- ++(0.125,0) -- ++(0,0.3) -- ++(0.125,0) -- ++(0,-0.3) -- ++(0.125,0) -- ++(0,0.3) -- ++(0.125,0) -- ++(0,-0.3) -- ++(0.125,0);
	}
}}

\node[draw,rectangle,align=center] (gate) {Gate\\[0]Flip-Flop};
\node[amp,left=6mm of gate,yshift= 5mm] (ampone) {};
\node[amp,left=6mm of gate,yshift=-5mm] (amptwo) {};
\node[block,right=5mm of gate] (and) {\&};
\node[block,right=5mm of and] (ctr) {Counter};
\node[clk,below=5mm of and] (clk) {};

\begin{pgfonlayer}{bg}
	\draw[->] (ampone.east) -- ++(-\pgflinewidth,0) -- ++(2.5mm,0) -- ++(0,-4mm) -- ++(4mm,0);
	\draw[->] (amptwo.east) -- ++(-\pgflinewidth,0) -- ++(2.5mm,0) -- ++(0, 4mm) -- ++(4mm,0);
	\draw[->] (clk.north) -- (and.south);
	\draw[->] (gate.east) -- (and.west);
	\draw[->] (and.east)  -- (ctr.west);
	\draw[-] (ampone.west) -- ++(\pgflinewidth,0) -- ++(-10mm,0) node [draw,circle,radius=1pt,inner sep=1pt,fill=white] (inone) {};
	\draw[-] (amptwo.west) -- ++(\pgflinewidth,0) -- ++(-10mm,0) node [draw,circle,radius=1pt,inner sep=1pt,fill=white] (intwo) {};
\end{pgfonlayer}

\node[anchor=east,xshift=-0.5ex] at (clk.west) {Digital clock};
\node[anchor=west,xshift=0.5ex]  at (clk.east) {$f_\text{TIC}$};
\node[anchor=north] at (inone) {$x_\text{pulse,1}(t)$};
\node[anchor=north] at (intwo) {$x_\text{pulse,2}(t)$};
\end{tikzpicture}

\vspace{-3mm}

\begin{tikzpicture}
\tikzstyle{every node}=[font=\small]
\end{tikzpicture}

\begin{tikzpicture}
\tikzstyle{every node}=[font=\small]
\tikzstyle{every path}=[thick]

\def\width{7}
\def\highlvl{0.3}
\def\highone{0.81}
\def\hightwo{5.06}
\def\highdur{0.4}

\begin{scope}[xshift=1cm]
	\draw[-,gray,dashed] (\highone,0.5) -- (\highone,-2.2);
	\draw[-,gray,dashed] (\hightwo,0.5) -- (\hightwo,-2.2);
\end{scope}

\node[yshift=0.5ex] (xpulse1) at (0,0) {$x_\text{pulse,1}(t)$};
\begin{scope}[xshift=1cm]
	\draw[-] (0,0) -- (\highone,0) -- (\highone,\highlvl) -- (\highone+\highdur,\highlvl) -- (\highone+\highdur,0) -- (\width,0);
\end{scope}

\node[yshift=0.5ex] (xpulse2) at (0,-0.5) {$x_\text{pulse,2}(t)$};
\begin{scope}[xshift=1cm,yshift=-0.5cm]
	\draw[-] (0,0) -- (\hightwo,0) -- (\hightwo,\highlvl) -- (\hightwo+\highdur,\highlvl) -- (\hightwo+\highdur,0) -- (\width,0);
\end{scope}

\node[yshift=0.5ex] (gate) at (0,-1) {Gate};
\begin{scope}[xshift=1cm,yshift=-1cm]
	\draw[-] (0,0) -- (\highone,0) -- (\highone,\highlvl) -- (\hightwo,\highlvl) -- (\hightwo,0) -- (\width,0);
\end{scope}

\node[yshift=0.5ex] (gate) at (0,-1.5) {Clock};
\begin{scope}[xshift=1cm,yshift=-1.5cm]
	\foreach \x in {0.0,0.25,...,6.75}
		\draw (\x,0) -- ++(0,\highlvl) -- ++(0.125,0) -- ++(0,-\highlvl) -- ++(0.125cm+0.5\pgflinewidth,0);
\end{scope}

\node[yshift=0.5ex] (gate) at (0,-2) {Counts};
\begin{scope}[xshift=1cm,yshift=-2cm]
	\draw[-] (0,0) -- (1cm+0.5\pgflinewidth,0);
	\foreach \x in {1.0,1.25,...,5.00}
		\draw (\x,0) -- ++(0,\highlvl) -- ++(0.125,0) -- ++(0,-\highlvl) -- ++(0.125cm+0.5\pgflinewidth,0);
	\draw[-] (5.25,0) -- (\width,0);
\end{scope}
\end{tikzpicture}

\vspace{-5mm}
\end{center}
\caption{Non-interpolating Time Interval Counter block diagram and signal timing graph.
Note that the TIC's precision depends on its digital clock frequency $ f_\text{TIC} $.}
\label{fig:tic}
\end{figure}

Figure~\ref{fig:tic} illustrates the operation of a non-interpolating TIC.
The time resolution can be significantly improved at the cost of increased system complexity by implementing an interpolating TIC \cite{szplet2014subpicosecond}.

\subsection{Dual Mixer Time Difference}
\label{sec:dmtd}

The original DMTD method utilizes an analog, discrete circuit to measure the phase difference $ \Delta \varphi (t) $ between two sine waves with (almost) identical frequency \cite{allan1975picosecond}.

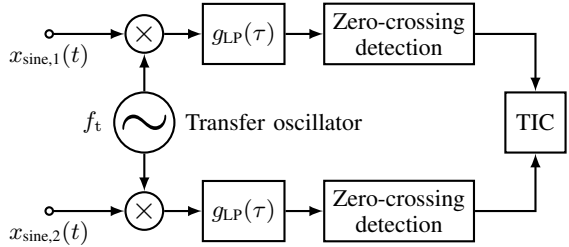
\begin{figure}[ht!]
\begin{center}
\vspace{-2mm}
\begin{tikzpicture}
\tikzset{block/.style={%
	rectangle, draw, minimum height=8mm, minimum width=8mm
}}
\tikzset{operand/.style={%
	circle, draw, minimum size=5mm, inner sep=0
}}
\tikzset{osc/.style={%
	circle, draw, minimum size=8mm, inner sep=0,
	path picture={%
		\node[yshift=-0.35ex] at (0,0) {\Huge $\sim$};
	}
}}

\tikzstyle{every node}=[font=\small]
\tikzstyle{every path}=[thick]

\node[osc] (osc) {};
\node[operand,above=5mm of osc] (mulone) {\large $\times$};
\node[operand,below=5mm of osc] (multwo) {\large $\times$};
\node[block,right=5mm of mulone] (lpone) {$ g_\text{LP}(\tau) $};
\node[block,right=5mm of multwo] (lptwo) {$ g_\text{LP}(\tau) $};
\node[block,right=5mm of lpone,align=center] (zcone) {Zero-crossing\\[0]detection};
\node[block,right=5mm of lptwo,align=center] (zctwo) {Zero-crossing\\[0]detection};
\node[block] (tic) at (52mm,0) {TIC};

\begin{pgfonlayer}{bg}
	\draw[->] (osc.north) -- (mulone.south);
	\draw[->] (osc.south) -- (multwo.north);
	\draw[->] (mulone.east) -- (lpone.west);
	\draw[->] (multwo.east) -- (lptwo.west);
	\draw[->] (lpone.east) -- (zcone.west);
	\draw[->] (lptwo.east) -- (zctwo.west);
	\draw[->] (zcone.east) -| (tic.north);
	\draw[->] (zctwo.east) -| (tic.south);
	\draw[<-] (mulone.west) -- ++(-10mm,0) node [draw,circle,radius=1pt,inner sep=1pt,fill=white] (inone) {};
	\draw[<-] (multwo.west) -- ++(-10mm,0) node [draw,circle,radius=1pt,inner sep=1pt,fill=white] (intwo) {};
\end{pgfonlayer}

\node[anchor=west]  at (osc.east) {Transfer oscillator};
\node[anchor=east] at (osc.west) {$f_\mathrm{t}$};
\node[anchor=north] at (inone) {$x_\text{sine,1}(t)$};
\node[anchor=north] at (intwo) {$x_\text{sine,2}(t)$};
\end{tikzpicture}
\vspace{-5mm}
\end{center}
\caption{Dual Mixer Time Difference block diagram.}
\label{fig:dmtd_block}
\end{figure}

Fundamentally, the time difference between the sine waves' zero-crossings is measured using a TIC.
However, instead of using the original reference signal at frequency $ f_\mathrm{r} $, which would require a highly accurate TIC with a high-frequency digital clock, the sine wave reference signal is down-converted to ease the constraints on the TIC.
Figure~\ref{fig:dmtd_block} shows a typical DMTD block diagram illustrating this principle.

Real-valued down-conversion with an analog mixing circuit and a transfer oscillator at frequency $ f_\mathrm{t} $ is employed to down-shift the reference signal from frequency $ f_\mathrm{r} $ to a very low intermediate frequency, the so-called beat frequency $ f_\mathrm{b} = \left| f_\mathrm{r} - f_\mathrm{t} \right| $.
A low-pass filter $ g_\mathrm{LP}(\tau) $ is applied to eliminate interfering signal components (image frequency and up-converted Direct Current (DC) bias) and to reduce the noise bandwidth, resulting in the beat signal:
\begin{alignat}{1}
\label{eqn:x_beat}
x_\text{beat}(t) & = \left\{ x_\text{sine}(t) \cdot \, \mathrm{sin}(2\pi f_\mathrm{t}t) \right\} \ast g_\mathrm{LP}(\tau) \\
                 & = \left\{ \mathrm{sin}(2 \pi f_\mathrm{r} t + \varphi_\mathrm{e}(t)) \cdot \, \mathrm{sin}(2\pi f_\mathrm{t}t) \right\} \ast g_\mathrm{LP}(\tau) \nonumber \\
                 & = \mathrm{sin}(2\pi f_\mathrm{b}t + \varphi_\mathrm{e}(t)) \nonumber
\end{alignat}

Its phase error term $ \varphi_\mathrm{e}(t) $ is identical to that of the original signal $ x_\text{sine}(t) $ prior to down-conversion (cf. Equation~\ref{eqn:x_sine}).

\begin{figure}[ht!]
\begin{center}
\vspace{-3mm}
\begin{tikzpicture}
\tikzstyle{every node}=[font=\small]

\begin{axis}[
		width=8.5cm,
		height=4.0cm,
		domain=0:2,
		samples=201,
		xmin=0,
		xmax=2,
		tick align=outside,
		xtick pos=left,
		ytick pos=bottom,
		grid=both,
		xtick={0, 0.5, 1, 1.5, 2},
		ytick={-1, 0, 1},
		xlabel=$ t \cdot f_\mathrm{b} $,
		ylabel=\footnotesize Normalized amplitude
	]
	\addplot[thick,red] {sin(deg(2*pi*x - 1./8*pi))};
	\addplot[thick,blue]{sin(deg(2*pi*x - 5./8*pi))};

	\pgfplotsinvokeforeach{0,2} {
		\node[anchor=north,xshift=-0.7ex,yshift=0.2ex] at (axis cs:0.5 * #1 + 3./16,0){\tiny $ \Delta t_\text{TIC}[#1] $};
		\draw[<->,thick](axis cs:0.5 * #1 + 1./16,0)--(axis cs:0.5 * #1 + 5./16,0);
	}
	\pgfplotsinvokeforeach{1,3} {
		\node[anchor=north,xshift=0.5ex,yshift=0.2ex] at (axis cs:0.5 * #1 + 3./16,0){\tiny $ \Delta t_\text{TIC}[#1] $};
		\draw[<->,thick](axis cs:0.5 * #1 + 1./16,0)--(axis cs:0.5 * #1 + 5./16,0);
	}
\end{axis}
\end{tikzpicture}
\vspace{-5mm}
\end{center}
\caption{Exemplary Dual Mixer Time Difference beat signals after down-conversion and low-pass filtering (cf. Equation~\ref{eqn:x_beat}).
Note the two zero-crossings per period and the time-difference between consecutive zero-crossings $ \Delta t_\text{TIC}[n] $, which is measured using a TIC.}
\label{fig:dmtd_beat}
\end{figure}
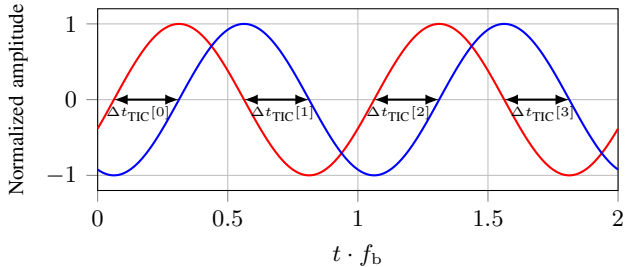

Figure~\ref{fig:dmtd_beat} illustrates the resulting beat signals, which will be fed to a TIC to gauge the time difference between their zero-crossings $ \Delta t_\text{TIC}[n] $.
The time error of the original signal can be computed as follows:
\[
\Delta t_\text{sine}[n] = \frac{f_\mathrm{r}}{f_\mathrm{b}} \cdot \Delta t_\text{TIC}[n]
\]

As can be concluded from Figure~\ref{fig:dmtd_beat}, the beat frequency determines the amount of zero-crossings and thus measurements per second ($ 2 f_\mathrm{b} $).
It also dictates the heterodyne factor~$ \frac{f_\mathrm{r}}{f_\mathrm{b}} $ that quantifies the resolution gain over the TIC's native resolution (see above Equation). \cite{riley2008handbook}

\FloatBarrier

\section{Digital Signal Processing Approach}
\label{sec:dsp}

\IEEEPARstart{T}{he} presented TIC and DMTD measurement techniques all rely on custom hardware setups and mixed-signal processing, requiring comprehensive know-how and integration effort to build a precise measurement device.

Instead of building custom measurement hardware, commercial measurement gear equipped with Analog-to-Digital Converters (ADC) to sample the reference signals and subsequent Digital Signal Processing (DSP) to perform the time-difference measurement could be used.
This novel idea was published only recently%
\footnote{
Mochizuki et al. published an ADC-based measurement system in 2007.
However, their implementation is still a custom-hardware mixed-signal DMTD system that uses an ADC instead of a TIC. \cite{mochizuki2007frequency}
}
by Sherman and Jördens \cite{sherman2016oscillator} as well as us \cite{andrich2017using}, with this paper being an extension of our initial contribution.

Both parties describe measurement systems that utilize inexpensive Software Defined Radios (SDR) as general-purpose sampling platforms with built-in DSP capabilities and extensive progammability.
However, the described algorithms are hardware-independent and can be implemented with any sampling platform that enables real-time or offline access to the sampled data.

\subsection{DSP for Sine Wave Reference Signals}
\label{sec:sine_dsp}

The sine wave DSP algorithm and its analog DMTD counterpart share the same core principle, namely the use of frequency down-conversion and subsequent phase measurement.
However, the DMTD relies upon real-valued down-conversion, requiring zero-crossing detection for phase-measurement.
Thereby, both measurement rate and resolution are restricted according to the chosen beat frequency~$ f_\mathrm{b} $ (see Section~\ref{sec:dmtd}).

This significant limitation is overcome by the DSP algorithm's use of complex down-conversion of the real-valued input signals (at sampling frequency $ f_\mathrm{s} $).
A Numerically Controlled Oscillator (NCO) at frequency~$ f_\mathrm{t} $ is utilized as transfer oscillator.
A low-pass filter $ g_\text{LP}[m] $ is required to suppress potentially interfering signal components, i.e. image and up-converted DC bias.

We suggest choosing $ f_\mathrm{t} = f_\mathrm{r} $, to result a beat frequency $ f_\mathrm{b} = 0 $.%
\footnote{Sherman and Jördens \cite{sherman2016oscillator} suggest a non-zero beat frequency $ f_\mathrm{b} $ without giving an explanation for this choice.
We assume this to be adopted from the real-valued processing performed by analog DMTD systems.
Although a small, non-zero $ f_\mathrm{b} $ shouldn't affect the measurement, $ f_\mathrm{b}  = 0 $ is the logical choice if complex-valued down-conversion is used.}
The measured signal's phase error $ \varphi_\mathrm{e}[n] $ is then equal to the phase angle of the complex signal.
Otherwise, the phase angle contribution of $ f_\mathrm{b} $ must be subtracted, which may be necessary, if the NCO is unable to exactly synthesize the required $ f_\mathrm{t} $.
See Equations \ref{eqn:z_sine} and \ref{eqn:z_sine_arg} for an analytic description.
\begin{alignat}{1}
\label{eqn:z_sine}
z_\text{sine}[n] & = \left\lbrace x_\text{sine}(f_\mathrm{s} n) \cdot \, \mathrm{e}^{-\mathrm{j} 2\pi \frac{f_\mathrm{t}}{f_\mathrm{s}}n} \right\rbrace \ast g_\mathrm{LP}[m] \\
                 & = \left\lbrace \mathrm{sin} \! \left( 2 \pi \frac{f_\mathrm{r}}{f_\mathrm{s}}n + \varphi_\mathrm{e}[n] \right) \cdot \, \mathrm{e}^{-\mathrm{j} 2\pi \frac{f_\mathrm{t}}{f_\mathrm{s}}n} \right\rbrace \ast g_\mathrm{LP}[m] \nonumber \\
                 & = \mathrm{e}^{\mathrm{j}2\pi \frac{f_\mathrm{b}}{f_\mathrm{s}}n} \cdot \mathrm{e}^{\mathrm{j} \varphi_\mathrm{e}[n]} \nonumber \\
\label{eqn:z_sine_arg}
\varphi_\mathrm{e}[n] & = \mathrm{arg} \! \left\lbrace z_\text{sine}[n] \right\rbrace - \mathrm{arg} \! \left\lbrace \mathrm{e}^{\mathrm{j}2\pi \frac{f_\mathrm{b}}{f_\mathrm{s}}n} \right\rbrace
\end{alignat}

The algorithm yields a valid $ \varphi_\mathrm{e}[n] $ measurement for every sample of the input data (band-limited by $ g_\mathrm{LP}[m] $).
Depending on the intended application, all output samples can be used, e.g. to compute the phase noise power-spectral density.
Alternatively, the sample rate can be decreased by means of decimation (i.e. low-pass filtering and down-sampling) to facilitate long-term measurements with enhanced precision due to reduced noise bandwidth.

\begin{figure}[ht!]
\begin{center}
\vspace{-3mm}
\begin{tikzpicture}
\tikzset{block/.style={%
	rectangle, draw, minimum height=8mm, minimum width=8mm
}}
\tikzset{operand/.style={%
	circle, draw, minimum size=5mm, inner sep=0
}}
\tikzset{osc/.style={%
	circle, draw, minimum size=8mm, inner sep=0,
	path picture={%
		\node[yshift=-0.35ex] at (0,0) {\Huge $\sim$};
	}
}}

\tikzstyle{every node}=[font=\small]
\tikzstyle{every path}=[thick]

\node[osc] (osc) {};
\node[operand,above=4mm of osc] (mulone) {\large $\times$};
\node[operand,below=4mm of osc] (multwo) {\large $\times$};
\node[block,right=5mm of mulone] (lpone) {$ g_\text{LP}[m] $};
\node[block,right=5mm of multwo] (lptwo) {$ g_\text{LP}[m] $};
\node[operand] (div) at (24.1mm,0) {\large $\div$};
\node[block,right=5mm of div] (arg) {$ \mathrm{arg}\{\cdot\} $};

\begin{pgfonlayer}{bg}
	\draw[<-] (mulone.west) -- ++(-10mm,0) node [draw,circle,radius=1pt,inner sep=1pt,fill=white] (inone) {};
	\draw[<-] (multwo.west) -- ++(-10mm,0) node [draw,circle,radius=1pt,inner sep=1pt,fill=white] (intwo) {};
	\draw[->,double,double distance=1\pgflinewidth] (osc.north) -- (mulone.south);
	\draw[->,double,double distance=1\pgflinewidth] (osc.south) -- (multwo.north);
	\draw[->,double,double distance=1\pgflinewidth] (mulone.east) -- (lpone.west);
	\draw[->,double,double distance=1\pgflinewidth] (multwo.east) -- (lptwo.west);
	\draw[->,double,double distance=1\pgflinewidth] (lpone.east) -| (div.north);
	\draw[->,double,double distance=1\pgflinewidth] (lptwo.east) -| (div.south);
	\draw[->,double,double distance=1\pgflinewidth] (div.east) -- (arg.west);
	\draw[-] (arg.east) -- ++(7mm,0) node [draw,circle,radius=1pt,inner sep=1pt,fill=white] (out) {};
\end{pgfonlayer}

\node[anchor=west]  at (osc.east) {NCO};
\node[anchor=east] at (osc.west) {$f_\mathrm{t}$};
\node[anchor=north] at (inone) {$x_\text{sine,1}[n]$};
\node[anchor=north] at (intwo) {$x_\text{sine,2}[n]$};
\node[anchor=north] at (out) {$\Delta\varphi[n]$};
\end{tikzpicture}
\vspace{-5mm}
\end{center}
\caption{Sine reference signal DSP algorithm block diagram.
Note that single lines denote real-valued signals and double lines indicate complex signals.}
\label{fig:dsp_sine}
\end{figure}
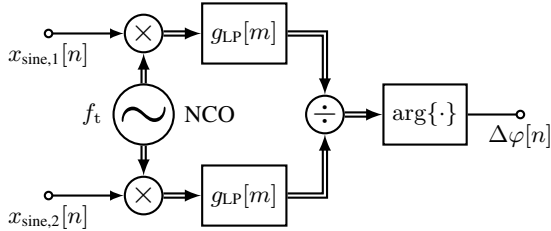

However, the computed phase error is only relative to the NCO that is implicitly driven by the sampling clock, which is inadequate to gauge high-performance reference oscillators.
To compare two reference sources against each other, their complex signals can be divided, yielding a complex residual signal.
See Figure~\ref{fig:dsp_sine} for an illustration of the algorithm.

The relevance of phase measurement is not limited to reference signals for synchronization.
The IEEE Standard C37.118.1-2011~\cite{IEEEstdSynchrophasor} addresses synchronous phasor measurements for power grids.
Its non-normative Annex~C defines a measurement algorithm, which is also based on complex down-conversion in the digital domain.

While synchronous phasor measurements must provide fast response times, our approach is tuned for high precision by applying rigorous baseband filtering with very high stopband attenuation.
Furthermore, our approach employs dual-signal comparison (cf.~Figure~\ref{fig:dsp_sine}), which can improve the measurement precision for 10\,MHz signals by up to an order of magnitude, as explained in \cite{sherman2016oscillator}.

\subsection{DSP for Pulse Reference Signals}
\label{sec:pulse_dsp}

As described in Section~\ref{sec:pulse_sig}, pulse reference signals are best used as trigger events, wherefore the signal amplitude must exceed a defined threshold.
To check for this with an ADC-based system, each sample's value could be compared against said threshold.
However, this na\"ive approach would only provide a time resolution limited to the inverse of the sample rate $ f_\mathrm{s} $.

A simple TIC as described in Section~\ref{sec:tic} only possesses binary high/low information of a pulse signal.
In contrast, a sampled version of the same pulse holds valuable amplitude information that can be used to interpolate the exact location of the signal edge exceeding the threshold level.

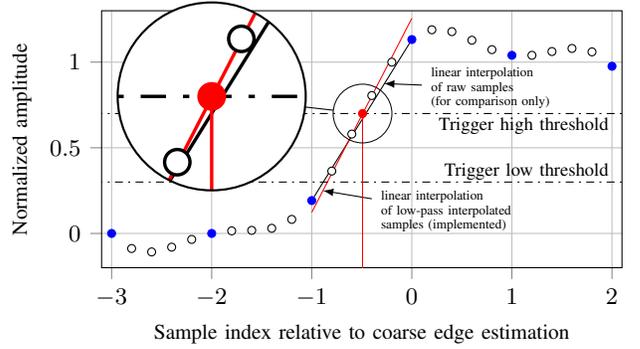
\begin{figure}[ht!]
\begin{center}
\vspace{-3mm}
\begin{tikzpicture}[spy using outlines={circle, magnification=3.2, connect spies}]
\tikzstyle{every node}=[font=\small]

\begin{axis}[
		width=8.5cm,
		height=5.0cm,
		xmin=-3.1,
		xmax=2.1,
		ymin=-0.2,
		ymax=1.3,
		grid=both,
		tick align = outside,
		xtick pos = left,
		ytick pos = bottom,
		xtick={-3, -2, -1, 0, 1, 2},
		ytick={0, 0.5, 1},
		xlabel=\footnotesize Sample index relative to coarse edge estimation,
		ylabel=\footnotesize Normalized amplitude
	]
	
	\addplot[
		only marks,
		color=blue,
		mark=*,
		mark options={scale=0.75}
	] coordinates {
		(-3.000, -0.000)
		(-2.000, -0.000)
		(-1.000, 0.192)
		(0.000, 1.132)
		(1.000, 1.040)
		(2.000, 0.976)
	};
	\addplot[
		only marks,
		color=black,
		mark=*,
		mark options={scale=0.75, fill=white}
	] coordinates {
		(-2.800, -0.088)
		(-2.600, -0.108)
		(-2.400, -0.080)
		(-2.200, -0.035)
		(-1.800, 0.015)
		(-1.600, 0.018)
		(-1.400, 0.030)
		(-1.200, 0.082)
		(-0.800, 0.364)
		(-0.600, 0.580)
		(-0.400, 0.805)
		(-0.200, 1.000)
		(0.200, 1.189)
		(0.400, 1.178)
		(0.600, 1.128)
		(0.800, 1.073)
		(1.200, 1.040)
		(1.400, 1.062)
		(1.600, 1.080)
		(1.800, 1.060)
		(2.200, 0.821)
		(2.400, 0.610)
		(2.600, 0.376)
		(2.800, 0.161)
	};

	\draw[black] (axis cs:-1,0.192) -- (axis cs:0,1.132);
	\node[anchor=west, align=left] (interp_lin) at (axis cs:0.1,0.85) {
		\tiny linear interpolation\\[-0.5em]
		\tiny of raw samples\\[-0.5em]
		\tiny (for comparison only)
	};
	\draw[->] (interp_lin) -- (axis cs:-0.28,0.875);
	
	\draw[red] (axis cs:-1,0.123) -- (axis cs:0,1.256);
	\node[anchor=west, align=left] (interp_both) at (axis cs:-0.4,0.13) {
		\tiny linear interpolation\\[-0.5em]
		\tiny of low-pass interpolated\\[-0.5em]
		\tiny samples (implemented)
	};
	\draw[->] (interp_both) -- (axis cs:-0.89,0.25);

	\draw[dash pattern=on .75pt off 2pt on 3pt off 2pt] (axis cs:-10,0.7)--(axis cs:10,0.7);
	\node[anchor=north east,xshift=.5ex,yshift=0.5ex] at (axis cs:2,0.7){\scriptsize Trigger high threshold};
	\draw[dash pattern=on .75pt off 2pt on 3pt off 2pt] (axis cs:-10,0.3)--(axis cs:10,0.3);
	\node[anchor=south east,xshift=.5ex,yshift=-0.6ex] at (axis cs:2,0.3){\scriptsize Trigger low threshold};
	
	\addplot[
		only marks,
		color=red,
		mark=*,
		mark options={scale=0.75}
	] coordinates {
		(-0.4926, 0.7)
	};
	
	\coordinate (spypoint) at (axis cs:-0.4926,0.7);
	\coordinate (magnifyglass) at (axis cs:-2,0.8);
	
	\draw[red] (-0.4926, 0.7) -- (-0.4926, -1);
\end{axis}
\spy[black, size=2.5cm] on (spypoint) in node[fill=white] at (magnifyglass);
\end{tikzpicture}
\vspace{-5mm}
\end{center}
\caption{Illustration of pulse signal edge estimation algorithm combining low-pass interpolation and linear interpolation.
Plot shows sampled edge synthesized using Equation~\ref{eqn:x_pulse} (\textcolor{blue}{blue}) and low-pass interpolated samples (black, hollow).
Note how the low-pass interpolation improves the accuracy of the subsequent linear interpolation.}
\label{fig:pulse_interp}
\end{figure}

We suggest the following estimation algorithm to determine the exact edge location (see Figure~\ref{fig:pulse_interp} for an illustration based on a signal synthesized according to Section~\ref{sec:pulse_sig}):
\begin{enumerate}
\item Coarsely locate the rising/falling edge by applying a Schmitt-Trigger%
\footnote{Due to hysteresis thresholding, the Schmitt-Trigger provides highly reliable edge detection even in the presence of noise.} to the input signal at native sample rate.
This provides an observation window with a configurable number of samples.
\item Increase the sample density within the observation window in order to approximate the edge location by low-pass interpolation (e.g. zero-padding the spectrum).
\item Ascertain the fractional sample index of the trigger event by linear interpolation.
\end{enumerate}

\FloatBarrier

\section{Software Defined Radio Solution}

\IEEEPARstart{S}{DR}s are mixed-signal measurement platforms typically used for RF applications.
They are highly configurable, programmable and usually offer DSP capability on a Field Programmable Gate Array (FPGA) out of the box.
Although most SDRs rely on local oscillators and analog up/down-conversion to achieve a wide range of selectable RF bands, some offer direct baseband access to their ADCs and/or Digital-to-Analog Converters (DAC).

\begin{figure}[ht!]
\begin{center}
\vspace{-3mm}
\begin{tikzpicture}
\tikzset{block/.style={%
	rectangle, draw, minimum height=8mm, minimum width=8mm
}}
\tikzset{operand/.style={%
	circle, draw, minimum size=5mm, inner sep=0
}}
\tikzset{adc/.style={
	rectangle, fill=none, draw=none, minimum width=10mm+\pgflinewidth, minimum height=8mm+\pgflinewidth, inner sep=0,
	path picture={%
		\draw[-] (-5mm,0mm) -- ++(4mm,4mm) -- ++(6mm,-0mm) -- ++(0mm,-8mm) -- ++(-6mm,0mm) -- ++(-4mm,+4mm) -- cycle;
		\node[anchor=center,xshift=0.5ex] at (0,0) {ADC};
	}
}}
\tikzset{clk/.style={
	circle, draw, minimum size=8mm, inner sep=0,
	path picture={%
		\draw (-0.30625,-0.15) -- ++(0.125,0) -- ++(0,0.3) -- ++(0.125,0) -- ++(0,-0.3) -- ++(0.125,0) -- ++(0,0.3) -- ++(0.125,0) -- ++(0,-0.3) -- ++(0.125,0);
	}
}}
\tikzset{osc/.style={%
	circle, draw, minimum size=8mm, inner sep=0,
	path picture={%
		\node[yshift=-0.35ex,xshift=-4.1mm] at (0,0) {\Huge $\sim$};
	}
}}

\tikzstyle{every node}=[font=\small]
\tikzstyle{every path}=[thick]

\node[clk] (clk) {};
\node[osc,right=5mm of clk] (osc) {};
\node[adc,above=4mm-0.5\pgflinewidth of clk,xshift=-2mm] (adcone) {};
\node[adc,below=4mm-0.5\pgflinewidth of clk,xshift=-2mm] (adctwo) {};
\node[operand,above=5.5mm of osc] (mulone) {\large $\times$};
\node[operand,below=5.5mm of osc] (multwo) {\large $\times$};
\node[block,right=5mm of mulone] (decone) {Decimator};
\node[block,right=5mm of multwo] (dectwo) {Decimator};

\begin{pgfonlayer}{bg}
	\draw[-] (adcone.west) -- ++(\pgflinewidth,0) -- ++(-7mm,0) node [draw,circle,radius=1pt,inner sep=1pt,fill=white] (inone) {};
	\draw[-] (adctwo.west) -- ++(\pgflinewidth,0) -- ++(-7mm,0) node [draw,circle,radius=1pt,inner sep=1pt,fill=white] (intwo) {};
	\draw[->] (adcone.east) -- ++(-\pgflinewidth,0) -- (mulone.west);
	\draw[->] (adctwo.east) -- ++(-\pgflinewidth,0) -- (multwo.west);
	\draw[->] (clk.north) -- ++(0, 4.2mm);
	\draw[->] (clk.south) -- ++(0,-4.2mm);
	\draw[->] (clk.east) -- (osc.west);
	\draw[->,double,double distance=1\pgflinewidth] (osc.north) -- (mulone.south);
	\draw[->,double,double distance=1\pgflinewidth] (osc.south) -- (multwo.north);
	\draw[->,double,double distance=1\pgflinewidth] (mulone.east) -- (decone.west);
	\draw[->,double,double distance=1\pgflinewidth] (multwo.east) -- (dectwo.west);
	\draw[-,double,double distance=1\pgflinewidth] (decone.east) -- ++(7mm,0) node [draw,circle,radius=1pt,inner sep=1pt,fill=white] (outone) {};
	\draw[-,double,double distance=1\pgflinewidth] (dectwo.east) -- ++(7mm,0) node [draw,circle,radius=1pt,inner sep=1pt,fill=white] (outtwo) {};
\end{pgfonlayer}

\node[anchor=east]  at (clk.west) {Clock};
\node[anchor=west]  at (osc.east) {NCO};
\node[anchor=south] at (mulone.north) {DDC};
\node[anchor=north] at (multwo.south) {DDC};
\node[anchor=north] at (inone) {$x_\mathrm{A}(t)$};
\node[anchor=north] at (intwo) {$x_\mathrm{B}(t)$};
\node[anchor=north] at (outone) {$z_\mathrm{A}[n]$};
\node[anchor=north] at (outtwo) {$z_\mathrm{B}[n]$};
\end{tikzpicture}
\vspace{-5mm}
\end{center}
\caption{Real-mode baseband DSP chain of typical direct-conversion SDR receiver.
Note that when processing complex input data, no separate DDC is performed.
The decimator block uses low-pass filtering and downsampling to adjust the output sample rate, as the ADCs' sample rate is fixed.
Single lines denote real-valued signals, whereas double lines indicate complex signals.}
\label{fig:dsp_chain}
\end{figure}
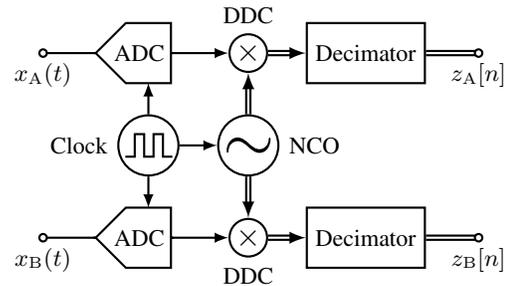

The baseband DSP of SDRs supports Digital Up/Down-Conversion (DUC/DDC) and decimators/interpolators to enable configurable sample rates.
Figure~\ref{fig:dsp_chain} illustrates a typical baseband DSP chain \cite{cruz2010sdr}.

To simultaneously measure both reference signal types, an SDR with 4 synchronous ADC channels that can be directly accessed is required.
Preferably, it also supports real-mode baseband DSP as depicted in Figure~\ref{fig:dsp_chain}.

Regardless of our choice for a specific SDR platform, we expect our suggested measurement technique to be applicable to a wide range of devices, not even limited to SDRs, since the algorithms described in Section~\ref{sec:dsp} are hardware-agnostic.
However, the widespread availability of SDRs and their built-in support for DSP primitives makes them a prime platform for implementation, enabling a comparatively cheap and easy to handle hardware solution.

\subsection{USRP Measurement Hardware}

The Universal Software Radio Peripheral (USRP) developed by Ettus Research LLC constitutes a flexible SDR platform, which is actively used within the academic community for a wide variety of communication or measurement scenarios.
The entry level USRP N210 costs less than \$2,000 and features a dual-channel ADC and a dual-channel DAC interfaced by an FPGA.
Although the N210 can be equipped with various RF front-ends, called daughterboards, the \$80 LFRX board enables direct baseband access to the dual-channel 14-bit ADC, also implementing the DSP chain illustrated in Figure~\ref{fig:dsp_chain}.

The DDC is implemented using a Coordinate Rotation Digital Computer (CORDIC) fixed-point algorithm \cite{volder1959cordic} and the decimator relies upon a Cascaded Integrator Comb (CIC) \cite{cic} filter with a final two-stage halfband Finite Impulse Response (FIR) filter \cite{doolittle2010halfband} to steepen the overall frequency response.
The combined filters enable integer decimation factors up to 500.

The N210 uses a Gigabit Ethernet connection to stream sampled data to a Personal Computer (PC) serving as a measurement host.
The link's maximum data rate is insufficient to transfer the ADC's 100 Mega Samples (MS) per second data stream, necessitating a minimal decimation factor of 4 (for 16-bit sample streaming) or 2 (for 8-bit sample streaming).

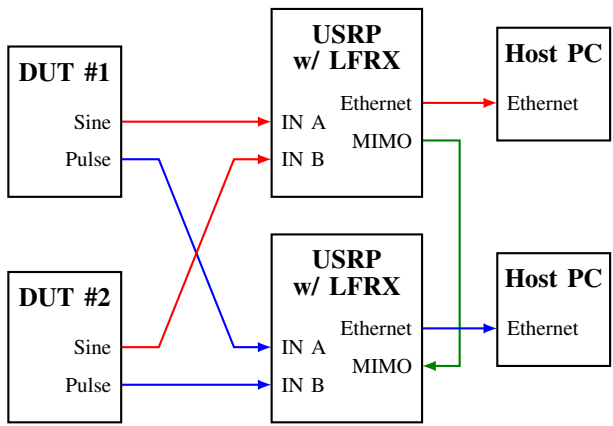
\begin{figure}[ht!]
\begin{center}
\vspace{-2mm}
\begin{tikzpicture}[
usrp/.pic={
	\draw (0,0) rectangle ++(20mm,25mm);
	\node[anchor=north,align=center,yshift=-0.5ex] (-label) at (10mm,25mm) {
		\normalsize \textbf{USRP}\\
		\normalsize \textbf{w/ LFRX}
	};
	\node[anchor=east] (-eth) at (20mm,12.5mm) {\footnotesize Ethernet};
	\node[anchor=west] (-inA) at (0mm,10.0mm) {\footnotesize IN A};
	\node[anchor=east] (-mimo) at (20mm,7.5mm) {\footnotesize MIMO};
	\node[anchor=west] (-inB) at (0mm,5mm) {\footnotesize IN B};
},
host/.pic={
	\draw (0,0) rectangle ++(15mm,15mm);
	\node[anchor=north,align=center,yshift=-0.5ex] (-label) at (7.5mm,15mm) {\normalsize \textbf{Host PC}};
	\node[anchor=west] (-eth) at (0mm,5mm) {\footnotesize Ethernet};
},
pics/dut/.style 2 args={code={
	\draw (0,0) rectangle ++(15mm,20mm);
	\node[anchor=north,align=center,yshift=-0.5ex] (-label) at (7.5mm,20mm) {
		\normalsize \textbf{#1}\\
		\normalsize \textbf{#2}
	};
	\node[anchor=east] (-sine) at (15mm,10mm) {\footnotesize Sine};
	\node[anchor=east] (-pulse) at (15mm,5mm) {\footnotesize Pulse};
}}]

\tikzstyle{every node}=[font=\small]
\tikzstyle{every path}=[thick]

\draw[thick] pic (dutone) at (0mm,30mm) {dut={DUT \#1}{}};
\draw[thick] pic (duttwo) at (0mm,0mm) {dut={DUT \#2}{}};

\draw[thick] pic (usrpone) at (35mm,30mm) {usrp};
\draw[thick] pic (usrptwo) at (35mm,0mm) {usrp};

\draw[thick] pic (hostone) at (65mm,37.5mm) {host};
\draw[thick] pic (hosttwo) at (65mm,7.5mm) {host};

\begin{pgfonlayer}{bg}
	\draw[->,red]  (dutone-sine.east)  -- (usrpone-inA.west);
	\draw[->,blue] (dutone-pulse.east) -- ++(5mm,0mm) -- ++(10mm,-25mm) -- (usrptwo-inA.west);
	\draw[->,red]  (duttwo-sine.east)  -- ++(5mm,0mm) -- ++(10mm,25mm) -- (usrpone-inB.west);
	\draw[->,blue] (duttwo-pulse.east) -- (usrptwo-inB.west);

	\draw[->,color=black!50!green] (usrpone-mimo.east) -- ++(5mm,0) -- ++(0,-30mm) -- (usrptwo-mimo.east);

	\draw[->,red]  (usrpone-eth.east) -- (hostone-eth.west);
	\draw[->,blue] (usrptwo-eth.east) -- (hosttwo-eth.west);
\end{pgfonlayer}
\end{tikzpicture}
\vspace{-3mm}
\end{center}
\caption{Schematic of the USRP hardware setup required to measure the sine and pulse signals of 2 Devices Under Test (DUT).
To evaluate only one signal type, a single, unsynchronized USRP is sufficient.
Actually, strict synchronicity is only required if a cross-device signal comparison is desired.}
\label{fig:setup_measure}
\end{figure}

A single, unsynchronized USRP N210 is sufficient to measure only one signal type.
However, if both sine and pulse signals are to be evaluated as depicted in Figure~\ref{fig:setup_measure}, a dual USRP setup must be employed to reach the required count of 4 coherently sampled channels.

We rely on a Multiple-Input Multiple-Output (MIMO) cable to achieve inter-device clock synchronization, but strict synchronicity is only required if signals are to be compared across the USRPs.
Both signals of each type are sampled on the same dual-channel ADC to improve sampling coherency.

\vspace{-1.em}
\subsection{Real-time Measurement Software}
\label{sec:software}

A real-time measurement software is implemented to significantly reduce the amount of data that need to be stored to disk compared to an off-line processing implementation.
The latter would require a 200\,MByte/s data stream to be stored, which would imply a tremendous storage volume for long-term measurements.

\vspace{.5em}
\subsubsection*{Sine Signal Processing}
A real-time implementation of the sine DSP algorithm described in Section~\ref{sec:sine_dsp} is supported by the USRP's DSP engine out of the box (compare Figures \ref{fig:dsp_sine} and \ref{fig:dsp_chain}).
Minimal programming effort on the host PC is required to configure the down-conversion and decimation DSP blocks appropriately.

Although the down-conversion and decimation filter algorithms (CORDIC and CIC) are implemented in 24-bit fixed-point arithmetic \cite{ettus:cordic_v}, the CIC filter offers only as little as 53\,dB of stopband attenuation \cite{cic}.
Therefore, we implemented single-precision floating-point down-conversion and FIR-based decimation on the host PC, so both DSP implementations can be directly compared (see Section~\ref{sec:cic_vs_fir}).
Additionally, the host decimator can be used in conjunction with the FPGA DSP, if a decimation factor higher than the latter's limit of 500 is desired.

The host DSP is implemented in the C~programming language and optimized by using the Single Instruction Multiple Data (SIMD) vector extensions of modern x86 processors.
This enables real-time processing with 25\,MS/s, providing better than 120\,dB stop band attenuation of the FIR low-pass filter.
Since 4 interleaved data streams must be processed in parallel (2 streams with an I/Q-pair each), vectorization can be realized efficiently.
The host FIR decimator supports compile-time configuration of stage count, decimation factors, and FIR coefficients, so the target sample rate can be chosen according to the use scenario.

For our measurements, a default of 3 stages with decimation factor 10 was chosen as a trade-off between resulting bandwidth (20\,kHz, 3\,dB) and data rate (0.4\,MByte/s), to enable long-term measurements without losing high-frequency phase noise information.

\vspace{.5em}
\subsubsection*{Pulse Signal Processing}
The pulse DSP algorithm from Section~\ref{sec:pulse_dsp} cannot be implemented on board the USRP without significant modification of its FPGA code.
To avoid time-consuming and intricate FPGA development, we implemented the algorithm on the host PC using double-precision floating-point arithmetic and a low-pass interpolation factor of 20.
No Single Instruction Multiple Data (SIMD) optimization effort was undertaken, since the interpolation is only performed for a limited number of samples, depending on the pulse signal's $ T_\text{period} $.
Finally, only the interpolation result is stored to disk, resulting in a negligible data rate and minimal storage volume.

\vspace{1em}
\subsubsection*{Software Implementation}

The USRP device family is supported by a wide variety of software development systems, including LabVIEW, MATLAB and GNU Radio.
However, all of them lack an important feature vital to synchronized multi-USRP measurements.
The User Datagram Protocol (UDP) used for sample transport between USRP and host PC is an unreliable transport layer protocol and is as such prone to packet loss.
If packet loss occurs, the affected samples are lost and leave a zero-length gap in the sample stream, causing multiple streams to be out-of-sync.

To mitigate packet loss, we use proprietary USRP software based on the USRP Hardware Driver (UHD) C++ library.
If packet loss is detected, the stream is zero-padded appropriately to ensure a gap-free sample stream.
Using a zero-copy ring buffer, this sample stream is accessed from multiple threads, distributing the load of sample reception and processing (i.e. computationally demanding FIR decimation).
This also relaxes the real-time processing requirements, because the ring buffer allows for a considerable data processing delay.

To further reduce the likelihood of packet loss, we employ a customized minimal Linux system with optimized network interface settings (large Direct Memory Access buffer and improved interrupt coalescing) that successfully demonstrated zero packet loss within a week of continuous measurement on a 2011 laptop PC\footnote{
The laptop form factor was chosen to build a compact measurement setup suitable for mobile use, e.g. outdoor GPSDO measurements.
The PC is equipped with an i7-620M processor (2.66\,GHz dual-core), 8 GiB random access memory, and a Gigabit Ethernet network interface.}.

\subsection{Post-processing Software}
\label{sec:post_proc}

As the real-time USRP software only performs limited pre-processing and data reduction, the actual evaluation is performed in a separate post-processing step.
This approach was chosen in order to evaluate recorded data in multiple ways.
The implemented post-processing focuses on the examined references' relative long-term drift.

First, the downmixed sine signal data is passed through additional high-attenuation FIR decimators to reduce its sample rate to 1~sample per second.
Subsequently, the phase difference between both sine signals is computed using complex division (cf. Figure~\ref{fig:dsp_sine}), unwrapped, and translated into a time error (see Equation~\ref{eqn:t_sine}).

The pulse signal pre-processing already produced interpolated edge times, so only the difference of both signals' rising edges is computed.

The Allan Variance is the standard statistical analysis method used in the field of frequency and time metrology \cite{std1139}.
Its aim is the description of random instabilities and the discrimination of the various modulated noise types that occur in oscillators (e.g. flicker and white noise) \cite{rutman1991characterization}.
Unfortunately, its design makes it immune to constant offset and linear drift, both of which are detrimental to the \textit{absolute} and \textit{relative} synchronization of measurement devices.
It is insufficient to use only the Allan Variance, if the synchronization-specific characteristics of an oscillator are to be considered.

Therefore, to determine the long-term drift of both signal types, a $1^\mathrm{st}$ order polynomial least-squares fit is applied.
The resulting estimated linear drift can be used to adjust most frequency standards that usually provide a mechanism to slightly tune the generated frequency.

Complementary to above linear drift estimation, we use Savitzky-Golay filters \cite{savgol}.
These apply least-squares, moving-window, arbitrary-order polynomial fitting to the input data, smoothing it according to the chosen window length and polynomial order.
Additionally, the same polynomial can be used to differentiate the (smoothed) input data, which we rely upon to ascertain the short-term drift of the time error within the configured observation window.
We have empirically chosen $2^\mathrm{nd}$ order polynomials to obtain a linear differentiation result.

\FloatBarrier

\section{Measurement Performance Evaluation}
\label{sec:perf_eval}

\IEEEPARstart{F}{or} the performance evaluation of our SDR measurement setup, we chose the most widely used reference signal types, i.e. 10\,MHz sine and 1\,PPS signals.
To determine the precision of such a system, generally the residual time-difference between two identical signals is measured, which are generated by power-splitting a single signal \cite{mochizuki2007frequency,sherman2016oscillator}.

\begin{figure}[ht!]
\begin{center}
\vspace{-2mm}
\begin{tikzpicture}[
usrp/.pic={
	\draw (0,0) rectangle ++(20mm,25mm);
	\node[anchor=north,align=center,yshift=-0.5ex] (-label) at (10mm,25mm) {
		\normalsize \textbf{USRP}\\
		\normalsize \textbf{w/ LFRX}
	};
	\node[anchor=east] (-eth) at (20mm,12.5mm) {\footnotesize Ethernet};
	\node[anchor=west] (-inA) at (0mm,10.0mm) {\footnotesize IN A};
	\node[anchor=east] (-mimo) at (20mm,7.5mm) {\footnotesize MIMO};
	\node[anchor=west] (-inB) at (0mm,5mm) {\footnotesize IN B};
},
host/.pic={
	\draw (0,0) rectangle ++(15mm,15mm);
	\node[anchor=north,align=center,yshift=-0.5ex] (-label) at (7.5mm,15mm) {\normalsize \textbf{Host PC}};
	\node[anchor=west] (-eth) at (0mm,5mm) {\footnotesize Ethernet};
},
pics/dut/.style 2 args={code={
	\draw (0,0) rectangle ++(15mm,22.5mm);
	\node[anchor=north,align=center,yshift=-0.5ex] (-label) at (7.5mm,22.5mm) {
		\normalsize \textbf{#1}\\
		\normalsize \textbf{#2}
	};
	\node[anchor=east] (-sine) at (15mm,10mm) {\footnotesize Sine};
	\node[anchor=east] (-pulse) at (15mm,5mm) {\footnotesize Pulse};
}}]

\tikzstyle{every node}=[font=\small]
\tikzstyle{every path}=[thick]

\draw[thick] pic (dut) at (5mm,15mm) {dut={SRS}{FS725}};

\draw[thick] pic (usrpone) at (35mm,30mm) {usrp};
\draw[thick] pic (usrptwo) at (35mm,0mm) {usrp};

\draw[thick] pic (hostone) at (65mm,37.5mm) {host};
\draw[thick] pic (hosttwo) at (65mm,7.5mm) {host};

\begin{pgfonlayer}{bg}
	\draw[<->,red]  (usrpone-inA.west) -- ++(-4mm,0) -- ++(0,-5mm) -- (usrpone-inB.west);
	\draw[<->,blue] (usrptwo-inA.west) -- ++(-4mm,0) -- ++(0,-5mm) -- (usrptwo-inB.west);
	\draw[-,red]  (dut-sine.east)  -- ++(5mm,0mm) -- ++(0mm,12.5mm)  -- ++(6mm,0)
		node [draw,circle,radius=1pt,inner sep=1pt,fill=red] {};
	\draw[-,blue] (dut-pulse.east) -- ++(5mm,0mm) -- ++(0mm,-12.5mm) -- ++(6mm,0)
		node [draw,circle,radius=1pt,inner sep=1pt,fill=blue] {};

	\draw[->,color=black!50!green] (usrpone-mimo.east) -- ++(5mm,0) -- ++(0,-30mm) -- (usrptwo-mimo.east);

	\draw[->,red]  (usrpone-eth.east) -- (hostone-eth.west);
	\draw[->,blue] (usrptwo-eth.east) -- (hosttwo-eth.west);
\end{pgfonlayer}
\end{tikzpicture}
\vspace{-3mm}
\end{center}
\caption{Reference measurement setup. By using tee-connectors, each USRP receives virtually identical input signals with zero drift. This is exploited to evaluate the setup's inherent measurement accuracy.}
\label{fig:setup_ref}
\end{figure}
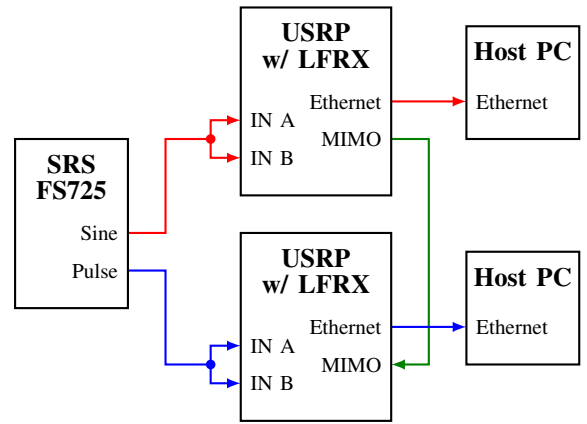

Figure~\ref{fig:setup_ref} shows the modified SDR setup used for the following measurements.
The inputs of the USRPs were kept at half-scale level to avoid distortion and to ensure a good Signal-to-Noise Ratio (SNR).

\vspace{-1em}

\FloatBarrier
\subsection{Comparison of USRP and Host-based DSP}
\label{sec:cic_vs_fir}

To compare the performance of both sine DSP implementations described in Section~\ref{sec:software}, we made multiple 1~hour measurements with different decimation factors $ n_\text{decim} $ and computed the SNR of the residual complex signal after division (cf. Figure~\ref{fig:dsp_sine}).

\begin{table}[ht!]
\begin{center}
\vspace{-.5em}
\caption{SNR-comparison of USRP-based and host-based 10\,MHz DSP}
\begin{tabular}{c|c|c|c}
\label{table:snr}
\textbf{\textit{n}$_\text{decim} $} & \textbf{SNR CIC (dB)} & \textbf{SNR FIR (dB)} & \textbf{$ \mathbf{\Delta}_\text{SNR} $ (dB)} \\ 
\hline 
10 & 69.3 & -- & -- \\ 
20 & 71.2 & 72.3 & 1.1 \\ 
40 & 73.4 & 74.7 & 1.3 \\ 
80 & 75.2 & 77.0 & 1.8 \\ 
100 & 75.9 & 77.7 & 1.8 \\ 
200 & 77.3 & 79.6 & 2.3 \\ 
400 & 78.0 & 81.1 & 3.1 \\ 
500 & 78.3 & 81.3 & 3.0 \\ 
\end{tabular}
\vspace{-.5em}
\end{center}
\end{table}

\begin{figure}[ht!]
\begin{center}
\begin{tikzpicture}
\tikzstyle{every node}=[font=\footnotesize]

\begin{loglogaxis}[
	width=84mm,
	height=40mm,
	xmin=1e-4,
	xmax=1e0,
	ymin=1e-12,
	ymax=1e-7,
	ytick={1e-7, 1e-8, 1e-9, 1e-10, 1e-11, 1e-12},
	grid=major,
	tick align=outside,
	xtick pos=left,
	ytick pos=bottom,
	xlabel={Averaging time $ \tau \; (\mathrm{s}) $},
	ylabel={\footnotesize Allan Deviation $ \sigma_\mathrm{y}(\tau) $},
	no markers,
	legend cell align=left,
	legend entries={USRP CIC, Host FIR},
]
	\addplot[thick,red,dash pattern=on 2pt off 1pt on 1pt off 1pt]
		 table {adev_cic500.csv};
	\addplot[thick,blue]
		table {adev_fir500.csv};
\end{loglogaxis}
\end{tikzpicture}
\vspace{-5mm}
\end{center}
\caption{Allan Deviation of FPGA-based CIC and host-based FIR decimators for $ n_\text{decim} = 500 $ during a 24~hour measurement.
The Allan Deviation is displayed up to $ \tau = 10^0 \, \mathrm{s} $ for better readability, but both curves' slope and difference continue until at least $ \tau = 10^4 \, \mathrm{s} $.
}
\label{fig:allan_dev_cic_vs_fir}
\end{figure}
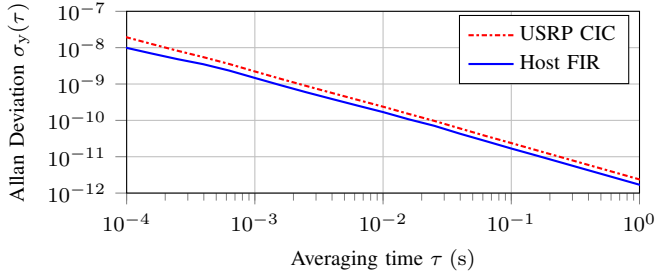

As can be seen from Table \ref{table:snr}, the host-based FIR decimation outperforms the FPGA-based alternative by up to 3\,dB for high decimation factors.
This is in good agreement with another 24~hour measurement that compares the Allan Deviation of both decimators for $ n_\text{decim} = 500 $ (see Figure~\ref{fig:allan_dev_cic_vs_fir}).

Consequentially, we suggest the use of host-based FIR decimation if maximum precision measurements are desired and its computational overhead is acceptable.
We use the FIR decimator for our subsequent measurements to benefit from its superior precision.

\subsection{Long-term Stability Measurement}

To evaluate the SDR measurement system's precision and long-term stability, a continuous 95-hour measurement was taken.
The resulting measurement data was evaluated as described in Section~\ref{sec:post_proc} with a 2 hour Savitzky-Golay window length.

\begin{figure}[ht!]
\begin{center}
\vspace{-4mm}
\includegraphics[scale=0.5]{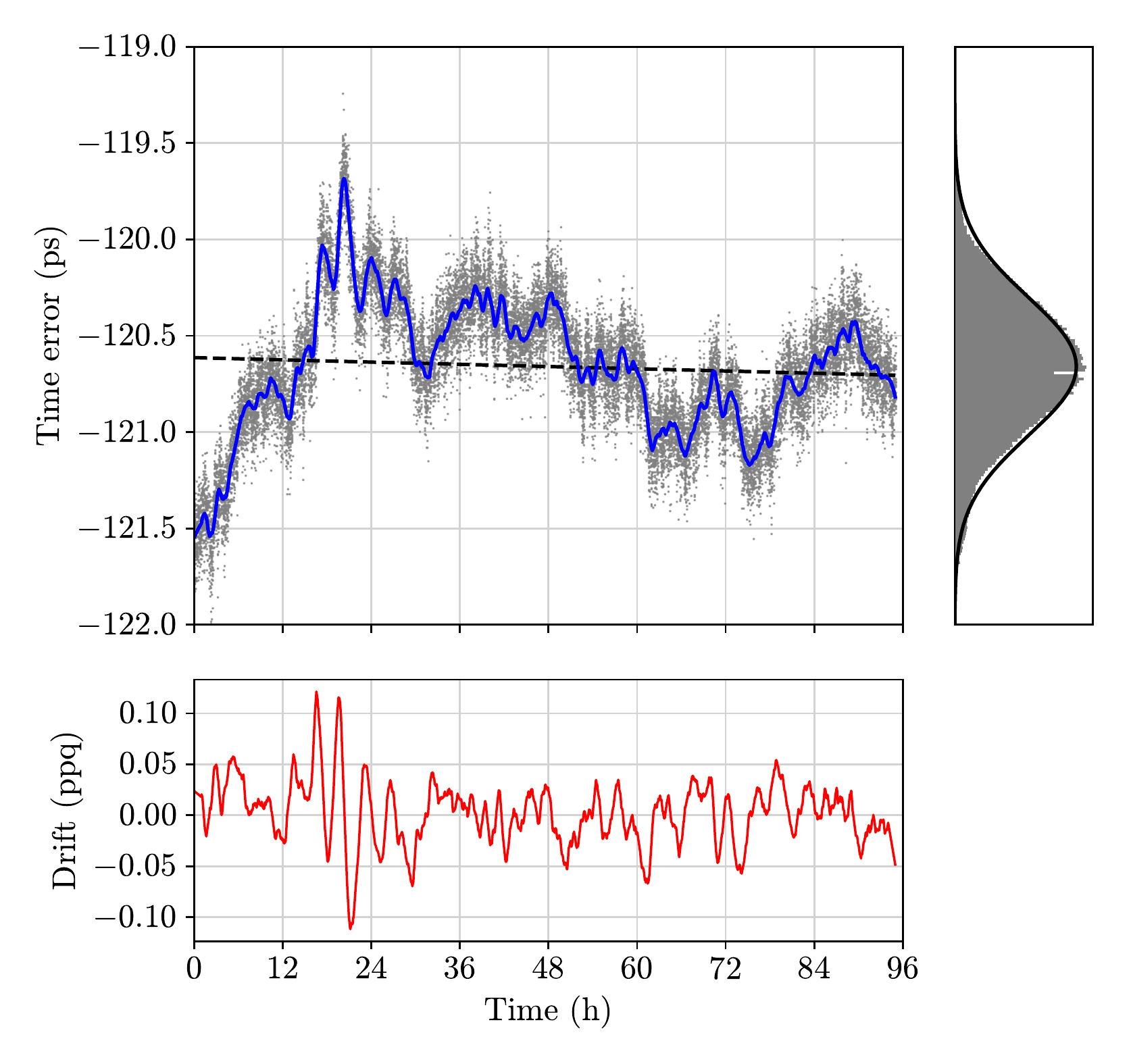}
\vspace{-5mm}
\end{center}
\caption{10\,MHz signal reference measurement results. Figure shows raw time error (gray dots), linear drift fit (black dashed line), raw time error histogram (gray), normal distribution (black), Savitzky-Golay filtered time error (\textcolor{blue}{blue}) and Savitzky-Golay differentiated short-time drift (\textcolor{red}{red})%
, the latter specified in parts-per-quadrillion (ppq, $10^{-15}$).
See Section~\ref{sec:post_proc} for time error and drift definitions.
Note the negligible linear drift ($-2.67 \cdot 10^{-19}$) and standard deviation ($ \sigma = 359\,\mathrm{fs} $).}
\label{plot:tee_10mhz_scatter}
\end{figure}

Figure~\ref{plot:tee_10mhz_scatter} illustrates the 10\,MHz signal measurement results.
Despite the raw data's erratic appearance, its histogram reveals a normal distribution.
The time error resolution is outstanding, with a $3\,\sigma$ accuracy of 1.08\,ps, which corresponds to a phase angle resolution of 14.0 seconds of arc.
These results can most likely be explained by the tremendous level of decimation, i.e. averaging, applied to the analog input signals.

\begin{figure}[ht!]
\begin{center}
\includegraphics[scale=0.5]{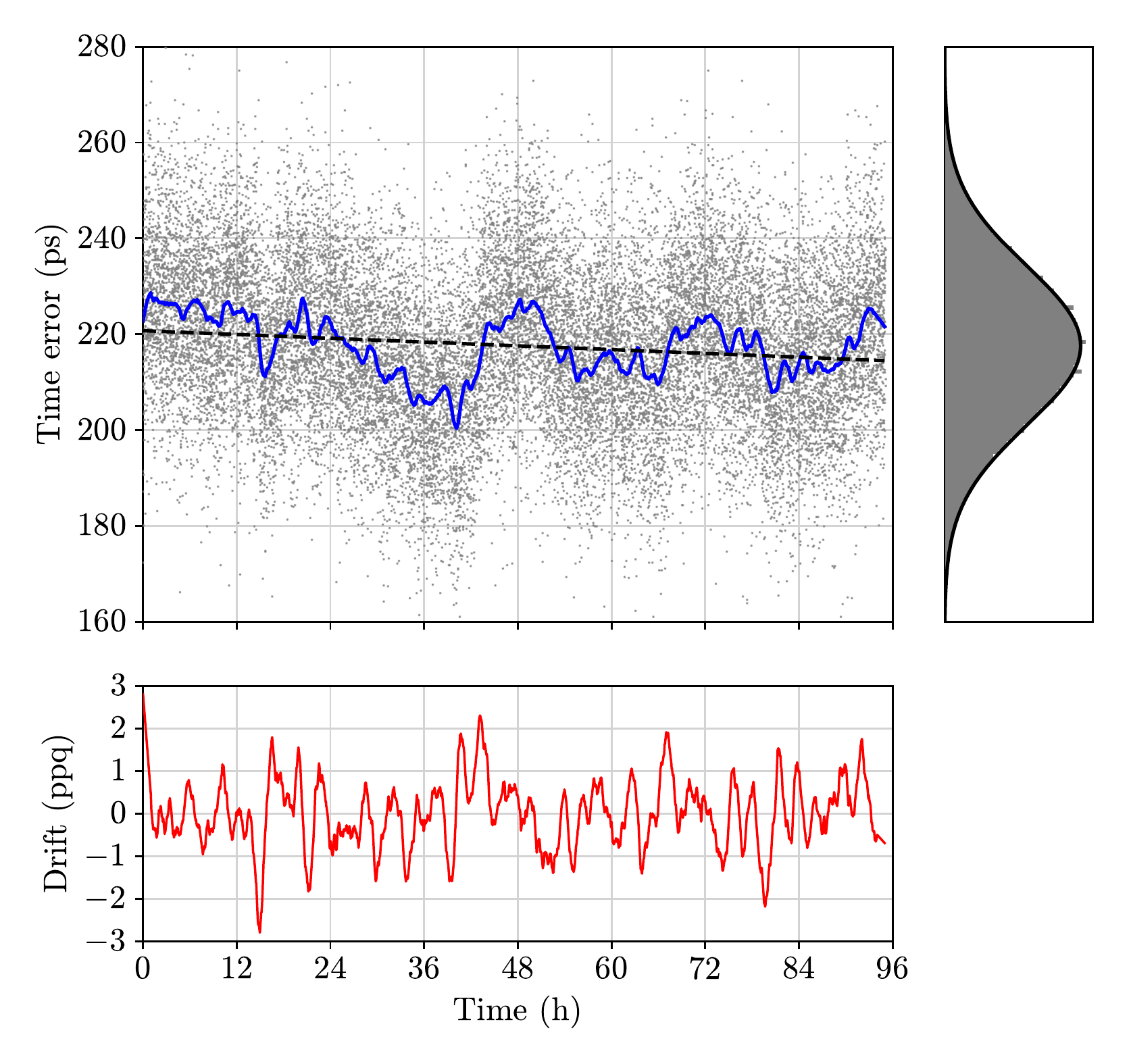}
\vspace{-5mm}
\end{center}
\caption{1\,PPS signal reference measurement results. Figure shows raw time error (gray dots), linear drift fit (black dashed line), raw time error histogram (gray), normal distribution (black), Savitzky-Golay filtered time error (\textcolor{blue}{blue}) and Savitzky-Golay differentiated short-time drift (\textcolor{red}{red})%
, the latter specified in parts-per-quadrillion (ppq, $10^{-15}$).
See Section~\ref{sec:post_proc} for time error and drift definitions.
Note the minimal linear drift ($-1.83 \cdot 10^{-17}$) and standard deviation ($ \sigma = 16.60\,\mathrm{ps} $).}
\label{plot:tee_1pps_scatter}
\end{figure}

Figure~\ref{plot:tee_1pps_scatter} depicts the 1\,PPS signal measurement results.
Although significantly worse than the 10\,MHz values, the $3\,\sigma$ time error resolution of 50\,ps is still excellent, considering the high level of interpolation applied to the 25\,MS/s input data (i.e. 40\,ns of sample spacing).

\begin{table}[ht!]
\begin{center}
\caption{10\,MHz and 1\,PPS reference measurement results}
\label{table:ref_results}
\begin{tabular}{r|c|c}
 & \textbf{10\,MHz} & \textbf{1\,PPS} \\ 
\hline 
$ \sigma $ & $ 359 $\,fs & $ 16.6 $\,ps \\ 
$ \mu $ & $ -121 $\,ps & $ 218 $\,ps \\ 
Linear drift & $ -2.67 \cdot 10^{-19} $ & $ -1.83 \cdot 10^{-17} $ \\ 
\end{tabular}
\end{center}
\end{table}

As can be concluded from Table \ref{table:ref_results}, both signal types' time errors show a comparatively high average value.
Since a constant difference is irrelevant for our intended drift measurements, we did not investigate possible sources.
However, different propagation delays are a likely explanation, since 120\,ps correspond to only 2.4\,cm of RG58 coax cabling.

If absolute measurement values are desired, further system calibration efforts beyond the scope of this paper become necessary.

\subsection{Comparison with DMTD, TIC, and SDR Systems}
\label{sec:comparison}

A direct comparison from a synchronization-centric point of view of DMTD and TIC systems with our SDR solution would be highly interesting.
However, we do not have DMTD or TIC hardware at our disposal, necessitating us to limit our comparison to published standard performance data, i.e. Allan Variance (DMTD) or standard deviation (TIC), respectively.
Regrettably, this rules out long-term stability considerations (cf. Figure~\ref{plot:tee_10mhz_scatter}, Figure~\ref{plot:tee_1pps_scatter}, and Section~\ref{sec:post_proc}).

\vspace{1em}
\subsubsection*{10\,MHz Signal}
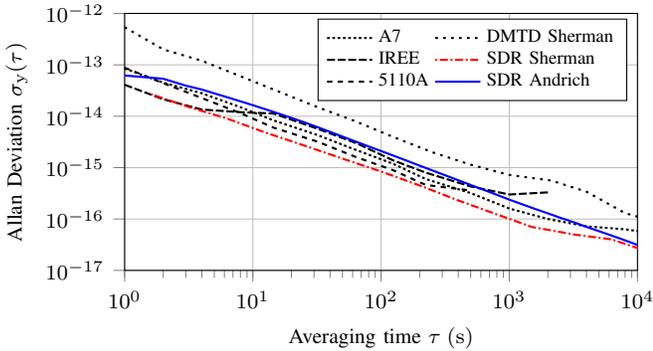
\begin{figure}[ht!]
\begin{center}
\vspace{-5mm}
\begin{tikzpicture}
\tikzstyle{every node}=[font=\footnotesize]

\begin{loglogaxis}[
	width=84mm,
	height=50mm,
	xmin=1e0,
	xmax=1e4,
	ymin=1e-17,
	ymax=1e-12,
	ytick={1e-12, 1e-13, 1e-14, 1e-15, 1e-16, 1e-17},
	grid=major,
	tick align=outside,
	xtick pos=left,
	ytick pos=bottom,
	xlabel={Averaging time $ \tau \; (\mathrm{s}) $},
	ylabel={\footnotesize Allan Deviation $ \sigma_\mathrm{y}(\tau) $},
	no markers,
	legend columns=2,
	legend cell align=left,
	legend style={font=\tiny, row sep=-2pt},
	legend entries={
		\scriptsize A7,
		\scriptsize DMTD Sherman,
		\scriptsize IREE,
		\scriptsize SDR Sherman,
		\scriptsize 5110A,
		\scriptsize SDR Andrich
	},
]
	\addplot[thick,black,dash pattern=on 1pt off 1pt]
		table {adev_dmtd_A7.csv};
	\addplot[thick,black,dash pattern=on 1pt off 2pt]
		table {adev_dmtd_sherman.csv};
	\addplot[thick,black,dash pattern=on 3pt off 1pt]
		table {adev_dmtd_IREE.csv};
	\addplot[thick,red,dash pattern=on .75pt off 1pt on 3pt off 1pt]
		table {adev_sdr_sherman.csv};
	\addplot[thick,black,dash pattern=on 2pt off 2pt]
		table {adev_dmtd_5110A.csv};
	\addplot[thick,blue]
		table {adev_sdr_andrich.csv};
\end{loglogaxis}
\end{tikzpicture}
\vspace{-5mm}
\end{center}
\caption{Allan Deviation of different 10\,MHz measurement systems, namely three DMTD systems (black lines, graphically extracted from  \cite{sojdr2003comparison}), an unknown DMTD system used for comparison by Sherman \cite{sherman2016oscillator}, and both SDR-based systems.
Note that the SDR solutions perform comparably to DMTD hardware.
Sherman's SDR results were post-processed using an averaging interval of $ \tau \approx 10^3\,\mathrm{s} $.
Therefore, their results are not comparable to ours (see Section~\ref{sec:comparison} for details).}
\label{plot:adev_comparison}
\end{figure}

Figure~\ref{plot:adev_comparison} uses the Allan Variance to compare the 10\,MHz signal measurement performance of multiple DMTD and both SDR systems.
Although our SDR measurement performs slightly worse than most analog DMTD systems for small $ \tau $, it outperforms all of them for $ \tau \gg 10^3 \, \mathrm{s} $.

The results of Sherman and Jördens are not directly comparable, because they employ a maximum likelihood time-domain optimization algorithm to determine the instantaneous frequency $ \dot{\varphi}_\mathrm{e}(t) $ (cf. Equation~\ref{eqn:x_sine}), assuming a linear phase drift during a variable observation window (see Appendix~A of their paper \cite{sherman2016oscillator}).
This approach reduces the signal's wideband noise and therefore improves its signal-to-noise ratio, decreasing the Allan Variance.

However, Sherman's algorithm will distort the spectral characteristics of the signal, if the instantaneous frequency is not constant within the entire window.
Particularly for long observation durations (Sherman chose $ \tau \approx 10^3\,\mathrm{s} $), a constant $ \dot{\varphi}_\mathrm{e}(t) $ cannot be unconditionally assumed.

We avoid any such distortions, by using only spectrum-preserving post-processing prior to the computation of the Allan Variance, i.e. low-pass decimation with relatively high cut-off frequencies ($ f_\text{c,3\,dB} \approx 0.8 \, f_\text{Nyquist} $).
Additionally, our investigation of decimation on the host PC using floating-point arithmetic  shows a 3\,dB precision gain over Sherman's use of the USRP's built-in fixed-point DSP (see Section~\ref{sec:cic_vs_fir}).
Therefore, we consider our algorithm a more general and more accurate approach.

\FloatBarrier

\vspace{1em}
\subsubsection*{1\,PPS Signal}
Table~\ref{table:tic_comparison} lists the standard deviations $ \sigma $ of multiple TIC implementations capable of 1\,PPS measurement, mostly taken from a comparison published by Szplet et al. \cite{szplet2016tic}.
The performance of our SDR system is comparable to that of FPGA-borne solutions (same order of magnitude).
However, it is entirely software-based and therefore avoids the integration effort required to build a measurement system using only FPGA hardware.
Since we are the first to publish such a solution \cite{andrich2017using}, there are no references with other SDR-based 1\,PPS measurements available for comparison.

\begin{table}[ht!]
\begin{center}
\caption{Measurement Performance of Time Interval Counters}
\label{table:tic_comparison}
\begin{tabular}{l|l|l|c}
\textbf{Ref., Year} & \textbf{Platform} & \textbf{Method} & \textbf{$ \bm{\sigma} $ (ps)} \\ 
\hline 
\cite{szplet2016tic} 2013 & FPGA & Equivalent coding line & 6 \\ 
\hline
\cite{szplet2014modular} 2014 & FPGA & Four-phase clock, & $150$ \\
& & Time coding line & \\ 
\hline
\cite{szplet2016tic} 2016 & FPGA & Multi-edge coding in & 6 \\ 
& & independent coding lines & \\
\hline
\cite{szplet2016tic} 2016 & FPGA & Equivalent coding line & 4.5 \\ 
\hline
This work & SDR & Low-pass and linear & 16.6 \\ 
& & interpolation & \\
\end{tabular}
\end{center}
\vspace{-1em}
\end{table}

\subsection{Comparison with Laboratory Measurement Equipment}

Instead of specialized DMTD or TIC gear, general purpose laboratory measurement hardware can be relied upon for reference signal analysis.
We have identified two devices to be used in a long-term stability comparison, namely a Rohde \& Schwarz FSUP Signal Source Analyzer and a Tektronix DPO7254 Digital Sampling Oscilloscope.
Similar devices should be available in any well-equipped RF laboratory, although results for other units will most likely vary.

\vspace{1em}
\subsubsection*{Signal Analyzer}

The Rohde \& Schwarz FSUP combines phase noise testing and spectrum analysis features in a single \$100,000~class device.
It supports various phase noise measurement scenarios, including the measurement of 2 Devices Under Test (DUT) against each other, which is required for a comparison with our SDR solution.
The double DUT measurement mandates control over at least one oscillator's frequency by supplying a tuning voltage, as is typical for Voltage Controlled Oscillators (VCO).
However, reference signal generators cannot be tuned like VCOs and the FSUP fails to lock its phase detector without tuning control.

Unfortunately, the FSUP does not enable the comparison of 2 reference signal generators as is supported by our SDR system.
Although the FSUP is solicited as phase noise tester, we assume that the use case was not envisioned or intended by the designers, which is unfortunate, considering the device's cost.
Consequently, the split-signal measurement to characterize the device's precision (cf. Section~\ref{sec:perf_eval}, Figure~\ref{fig:setup_ref}) is not possible, either.

\vspace{1em}
\subsubsection*{Digital Sampling Oscilloscope}

The Tektronix DPO7254 is a 4~channel, 40~Giga Samples (GS) per second Digital Sampling Oscilloscope (DSO) with 2.5\,GHz of analog bandwidth.
It is a \$30,000~class device with various built-in measurement features, including those required to measure the phase difference of sinusoidal signals and the time difference (i.e. delay) between the edges of pulse signals.
Thus, it supports the features required for a comparison with our SDR system out of the box.

We realized the split-signal measurement from Figure~\ref{fig:setup_ref}, using a single DPO7254 instead of 2 synchronized USRP N210.
The DPO's channels used for 1\,PPS measurement operated at the full 2.5\,GHz bandwidth for maximum time resolution, while the 10\,MHz measurement channels were band-limited at 20\,MHz to suppress wide-band noise.
In order to gain an insight into the DPO's measurement abilities, we compared 3 different configurations for 96~hours, each:
\begin{itemize}
\item Simultaneous measurement of \textbf{10\,MHz} and \textbf{1\,PPS} signals as enabled by our SDR setup.
Full quad-channel sample rate of \textbf{10\,GS/s} and 500\,ns observation window with 40-fold interpolation, as trade-off between both signal types' precision.
\item Only the \textbf{10\,MHz} signal, but at \textbf{20\,GS/s} dual-channel sample rate.
Peak observation window length of 50~$\mu$s without interpolation to minimize measurement dead-time.
\item Only the \textbf{1\,PPS} signal, but at \textbf{20\,GS/s} dual-channel sample rate.
Highest possible interpolation factor of 200 with 50\,ns observation window length to maximize time resolution.
\end{itemize}

\noindent We employed the built-in phase and/or delay measurement features of the DPO7254 and read out the values over Ethernet using the VXI-11 protocol, yielding exactly one measurement per second by using the 1\,PPS signal to trigger the DSO%
\footnote{Note that although the DPO7254 supports multiple measurements per second, it only refreshes its value display twice in that interval.
Curiously, the same limitation applies to the VXI-11 value readout.
Therefore, we did not attempt to increase the sample rate for the dual-channel 10\,MHz measurement by using an external trigger signal faster than the 1\,PPS.
For both other measurements, the 1\,PPS signal must be used as trigger, anyway.}.
Table~\ref{table:ref_dpo_comparison} lists the standard deviation $ \sigma $, mean $ \mu $, and linear drift of the 3 measurements outlined above, as well as our USRP results from Table~\ref{table:ref_results} for comparison.

Because different cables were used for both measurements (the DPO7254 has BNC inputs; the USRP N210 has SMA ports) the different $ \mu $ values are non-significant and subject to calibration, anyway.
All linear drifts are smaller than the corresponding $ \sigma $ values and therefore negligible.

\begin{table}[ht!]
\begin{center}
\caption{Performance comparison between SDR and DSO}
\label{table:ref_dpo_comparison}
\begin{tabular}{l|r|c|c}
 &  & \textbf{10\,MHz} & \textbf{1\,PPS} \\ 
\hline 
\textbf{USRP N210} & $ \sigma $ & $ 359 $\,fs & $ 16.6 $\,ps \\ 
4 channel & $ \mu $ & $ -121 $\,ps & $ 218 $\,ps \\ 
25\,MS/s & Linear drift & $ -2.67 \cdot 10^{-19} $ & $ -1.83 \cdot 10^{-17} $ \\ 
\hline
\textbf{DPO7254} & $ \sigma $ & $ 128 $\,ps & $ 30.5 $\,ps \\
4 channel & $ \mu $ & $120$\,ps & $ -1.75 $\,ps \\
10\,GS/s & Linear drift & $ -8.89 \cdot 10^{-17} $ & $ -3.70 \cdot 10^{-19} $ \\
\hline
\textbf{DPO7254} & $ \sigma $ & $107$\,ps & $32.1$\,ps \\
2 channel & $ \mu $ & $44.7$\,ps & $-8.60$\,ps \\
20\,GS/s & Linear drift & $ 4.93 \cdot 10^{-17} $ & $-1.87 \cdot 10^{-18}$ \\
\end{tabular}
\end{center}
\end{table}

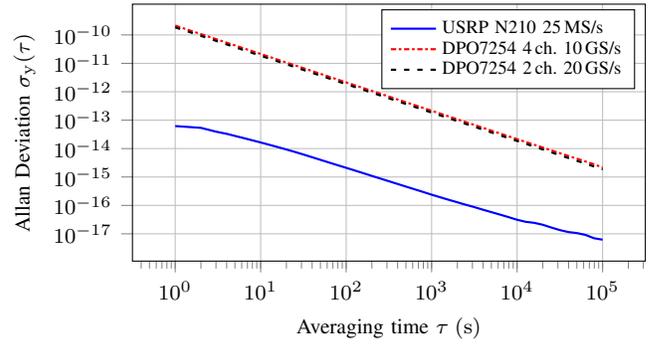
\begin{figure}[ht!]
\begin{center}
\vspace{-2mm}
\begin{tikzpicture}
\tikzstyle{every node}=[font=\footnotesize]

\begin{loglogaxis}[
	width=84mm,
	height=50mm,
	ytick={1e-10, 1e-11, 1e-12, 1e-13, 1e-14, 1e-15, 1e-16, 1e-17},
	grid=major,
	tick align=outside,
	xtick pos=left,
	ytick pos=bottom,
	xlabel={Averaging time $ \tau \; (\mathrm{s}) $},
	ylabel={\footnotesize Allan Deviation $ \sigma_\mathrm{y}(\tau) $},
	no markers,
	legend columns=1,
	legend cell align=left,
	legend style={row sep=-2.5pt},
	legend entries={
		\scriptsize USRP N210 25\,MS/s,
		\scriptsize DPO7254 4\,ch. 10\,GS/s,
		\scriptsize DPO7254 2\,ch. 20\,GS/s
	},
]
	\addplot[thick,blue]
		table {adev_n210_10mhz.csv};
	\addplot[thick,red,dash pattern=on 2pt off 1pt on 1pt off 1pt]
		table {adev_dpo_10mhz.csv};
	\addplot[thick,black,dash pattern=on 2pt off 3pt]
		table {adev_dpo_10mhz_2ch.csv};
\end{loglogaxis}
\end{tikzpicture}
\vspace{-5mm}
\end{center}
\caption{Allan Deviation of 10\,MHz signal long-term measurements.
The USRP outperforms the DPO by more than 2 orders of magnitude.
Note that the dual-channel DPO measurement enables a slight improvement over its quad-channel counterpart.}
\label{plot:adev_10mhz_dpo}
\end{figure}

As an additional statistical comparison of the 10\,MHz measurements, Figure~\ref{plot:adev_10mhz_dpo} illustrates their Allan Variances.
Similarly to the $ \sigma $ values, the USRP measurement setup outperforms the DPO by more than 2 orders of magnitude.
Although the DPO uses 400 (or even 800) times the sample rate of the USRP, this advantage is nullified by the DPO's substantial measurement dead-time (zero vs. $ \geq 99.995\,\% $) and higher noise bandwidth (0.8\,Hz vs. 20\,MHz), compared to the USRP.

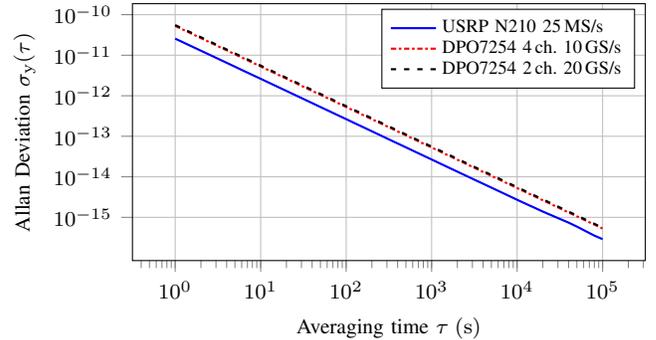
\begin{figure}[ht!]
\begin{center}
\vspace{-2mm}
\begin{tikzpicture}
\tikzstyle{every node}=[font=\footnotesize]

\begin{loglogaxis}[
	width=84mm,
	height=50mm,
	ytick={1e-10, 1e-11, 1e-12, 1e-13, 1e-14, 1e-15},
	grid=major,
	tick align=outside,
	xtick pos=left,
	ytick pos=bottom,
	xlabel={Averaging time $ \tau \; (\mathrm{s}) $},
	ylabel={\footnotesize Allan Deviation $ \sigma_\mathrm{y}(\tau) $},
	no markers,
	legend columns=1,
	legend cell align=left,
	legend style={row sep=-2.5pt},
	legend entries={
		\scriptsize USRP N210 25\,MS/s,
		\scriptsize DPO7254 4\,ch. 10\,GS/s,
		\scriptsize DPO7254 2\,ch. 20\,GS/s
	},
]
	\addplot[thick,blue]
		table {adev_n210_1pps.csv};
	\addplot[thick,red,dash pattern=on 2pt off 1pt on 1pt off 1pt]
		table {adev_dpo_1pps.csv};
	\addplot[thick,black,dash pattern=on 2pt off 3pt]
		table {adev_dpo_1pps_2ch.csv};
\end{loglogaxis}
\end{tikzpicture}
\vspace{-5mm}
\end{center}
\caption{Allan Deviation of 1\,PPS signal long-term measurements.
Note that the USRP outperforms the DPO by 3\,dB, despite its inferior analog bandwidth (i.e. time resolution) and sample rate, compared to the latter.}
\label{plot:adev_1pps_dpo}
\vspace{-4mm}
\end{figure}

Although the Allan Variation is usually relied upon to characterize sinusoidal oscillators, we also use it to statistically compare the 1\,PPS measurements (see Figure~\ref{plot:adev_1pps_dpo}).
Here, the USRP and the DPO are much closer, but the USRP still leads by a factor of 2.
Curiously, the DPO's precision deteriorates when switching from 10\,GS/s to 20\,GS/s.
This could be caused by the DPO internally interleaving 2 ADCs to double the sample rate at the potential cost of increased noise and jitter.

It seems quite impressive that an inexpensive SDR outperforms a DSO, which provides more than 200 times the analog bandwidth (i.e. time resolution) and 400 (or even 800) times the sample rate.
This is facilitated by the versatility of the Software Defined Radio platform, allowing to implement customized algorithms that achieve vast precision gains.
It also proves that our algorithm's combination of low-pass and linear interpolation (cf. Section~\ref{sec:pulse_dsp}) is well suited for the precise estimation of signal edges.

\FloatBarrier

\section{Conclusions}

\IEEEPARstart{W}{e} presented a system for high-precision measurement and analysis of reference signals, based on commercial off-the-shelf Software Defined Radio (SDR) hardware.
As SDR platforms are quite common in research laboratories, it seems likely to use them also for instrumentation in a cost-efficient way.
In this paper, we described how to analyze the stability of sine and pulse reference signals on the basis of the Universal Software Radio Peripheral (USRP) N210 by Ettus Research LLC.

Sine reference signals were digitally down-converted to baseband for the analysis of phase deviations.
For this purpose, out-of-the-box algorithms natively provided by common SDRs can be used as a basis.
As an extension, we demonstrated that the precision of the Digital Signal Processing chain (which is normally provided by SDRs using fixed-point arithmetic) can be improved by custom software, running on the host PC.
Using floating-point arithmetic in connection with Single Instruction Multiple Data (SIMD) vector extensions of modern processors, real-time performance can be provided even on moderate hardware.

The analysis of pulse reference signals was realized by a software trigger running on the host PC that precisely locates the trigger event time, i.e. where the slope passes a certain threshold.
Clearly, due to limited sample rate an adequate post-processing was needed.
Therefore, low-pass interpolation for increasing the sample density was applied, followed by a linear interpolation to ascertain the fractional sample index of the trigger event.

Using the USRP N210 SDR-platform, a measurement precision (standard deviation) of 0.36\,ps and 16.6\,ps was obtained for 10\,MHz and 1\,PPS reference signals, respectively, which is in the same order of magnitude compared to state-of-the-art methods such as the Dual Mixer Time Difference (DMTD) method and the Time Interval Counter (TIC).

Furthermore, joint and individual measurements of 10\,MHz and 1\,PPS signals with a Digital Sampling Oscilloscope (Tektronix DPO7254) showed that our proposed inexpensive SDR-based approach outperforms high-grade laboratory equipment by more than 2 orders of magnitude for 10\,MHz phase estimation and by factor 2 for 1\,PPS edge detection.
The SDR is also less bulky, cheaper, and more power efficient.

The proposed SDR-setup can be used out of the box, e.g. to check the integrity or rather for calibration of reference sources against each other.
It is suitable even for long-term measurements (days or weeks) and inherently provides a consistent storage of data.
A similar SDR-based approach was recently proposed by Sherman and Jördens~\cite{sherman2016oscillator} for the analysis of sine signals, only.
However, such an SDR-based system being capable of joint sine and pulse reference signal analysis is not found in literature, so far.

\FloatBarrier

\section*{Acknowledgements}

This work was supported by the Free State of Thuringia and the European Social Fund.

\IEEEtriggeratref{3}

\bibliographystyle{IEEEtran}
\bibliography{tim}

\end{document}